\input harvmac
\input epsf
\input amssym

\baselineskip 13pt



\newcount\figno
\figno=0
\def\fig#1#2#3{
\par\begingroup\parindent=0pt\leftskip=1cm\rightskip=1cm\parindent=0pt
\baselineskip=11pt
\global\advance\figno by 1
\midinsert
\epsfxsize=#3
\centerline{\epsfbox{#2}}
\vskip -21pt
{\bf Fig.\ \the\figno: } #1\par
\endinsert\endgroup\par
}
\def\figlabel#1{\xdef#1{\the\figno}}
\def\encadremath#1{\vbox{\hrule\hbox{\vrule\kern8pt\vbox{\kern8pt
\hbox{$\displaystyle #1$}\kern8pt}
\kern8pt\vrule}\hrule}}

\def\p{\partial}
\def\Ab{\overline{A}}
\def\Lamb{\overline{\Lambda}}
\def\zb{\overline{z}}

\def\taub{\overline{\tau}}
\def\Qc{{\cal Q}}
\def\Qcb{{\overline{\cal Q}}}
\def\hb{\overline{h}}
\def\ab{\overline{a}}
\def\lamb{\overline{\lambda}}
\def\fbar{\overline{f}}
\def\Pc{{\cal P}}
\def\eps{\epsilon}
\def\yb{\overline{y}}


\lref\Wald{
 R.M.~Wald,
  ``Black Hole Entropy is Noether Charge,''
Phys. Rev. D. 48 ,R3427. arXiv:9307038 [gr-qc].
}

\lref\Witten{
 Witten, Edward. ``2+ 1 dimensional gravity as an exactly soluble system." Nuclear Physics B 311.1 (1988): 46-78.
}

\lref\IyerWald{
 Iyer, Vivek, and Robert M. Wald. ``Some properties of the Noether charge and a proposal for dynamical black hole entropy." Physical review D 50.2 (1994): 846.
}

\lref\GutperleKraus{
Gutperle, Michael, and Per Kraus.``Higher spin black holes." Journal of High Energy Physics 2011.5 (2011): 1-24.
}

\lref\AmmonPerlmutterKraus{
Ammon, Martin, Per Kraus, and Eric Perlmutter. ``Scalar fields and three-point functions in D= 3 higher spin gravity." Journal of High Energy Physics 2012.7 (2012): 1-41.
}

\lref\AmmonCastroIqbal{
Ammon, Martin, Alejandra Castro, and Nabil Iqbal. ``Wilson lines and entanglement entropy in higher spin gravity." Journal of High Energy Physics 2013.10 (2013): 1-59.
}

\lref\BoerJottarB{
de Boer, Jan, and Juan I. Jottar. ``Entanglement Entropy and Higher Spin Holography in AdS $ _3$." arXiv preprint arXiv:1306.4347 (2013).
}

\lref\FaulknerGfC{
Faulkner, Thomas, et al. ``Gravitation from Entanglement in Holographic CFTs." arXiv preprint arXiv:1312.7856 (2013).
}

\lref\LashkariMcDermottRaamsdonk{
Lashkari, Nima, Michael B. McDermott, and Mark Van Raamsdonk. ``Gravitational Dynamics From Entanglement Thermodynamics". arXiv preprint arXiv:1308.3716 (2013).
}

\lref\Bekenstein{
Bekenstein, Jacob D. ``Black holes and the second law." Lettere Al Nuovo Cimento (1971–1985) 4.15 (1972): 737-740.
}

\lref\Banados{
Banados, Maximo, Claudio Teitelboim, and Jorge Zanelli. ``Black hole in three-dimensional spacetime." Physical Review Letters 69.13 (1992): 1849.
}

\lref\RyuTakayanagi{
Ryu, Shinsei, and Tadashi Takayanagi. ``Holographic derivation of entanglement entropy from the anti–de sitter space/conformal field theory correspondence." Physical review letters 96.18 (2006): 181602.
}

\lref\CasiniHuertaMyers{
Casini, Horacio, Marina Huerta, and Robert C. Myers. ``Towards a derivation of holographic entanglement entropy." Journal of High Energy Physics 2011.5 (2011): 1-41.
}

\lref\CompereSong{
Compère, Geoffrey, Juan I. Jottar, and Wei Song. ``Observables and microscopic entropy of higher spin black holes." arXiv preprint arXiv:1308.2175 (2013).
}

\lref\Cardy{
Cardy, John L. ``Operator content of two-dimensional conformally invariant theories." Nuclear Physics B 270 (1986): 186-204.
}

\lref\CardyCalabrese{
Calabrese, Pasquale, and John Cardy. ``Entanglement entropy and quantum field theory." Journal of Statistical Mechanics: Theory and Experiment 2004.06 (2004): P06002.
}

\lref\CastroGutperle{
Castro, Alejandra, et al. ``Conical defects in higher spin theories." Journal of High Energy Physics 2012.2 (2012): 1-34.
}

\lref\CampoleoniSSS{
Campoleoni, Andrea, et al. "Asymptotic symmetries of three-dimensional gravity coupled to higher-spin fields." Journal of High Energy Physics 2010.11 (2010): 1-36.
}

\lref\FaulknerRn{
Faulkner, Thomas. "The entanglement Renyi entropies of disjoint intervals in AdS/CFT." arXiv preprint arXiv:1303.7221 (2013).
}


\lref\AmmonNK{
  M.~Ammon, M.~Gutperle, P.~Kraus and E.~Perlmutter,
  ``Spacetime Geometry in Higher Spin Gravity,''
JHEP {\bf 1110}, 053 (2011).
[arXiv:1106.4788 [hep-th]].
}

\lref\CastroFM{
  A.~Castro, E.~Hijano, A.~Lepage-Jutier and A.~Maloney,
  ``Black Holes and Singularity Resolution in Higher Spin Gravity,''
JHEP {\bf 1201}, 031 (2012).
[arXiv:1110.4117 [hep-th]].
}

\lref\TanTJ{
  H.~-S.~Tan,
  ``Aspects of Three-dimensional Spin-4 Gravity,''
JHEP {\bf 1202}, 035 (2012).
[arXiv:1111.2834 [hep-th]].
}

\lref\HartmanGaberdiel{
Gaberdiel, Matthias R., Thomas Hartman, and Kewang Jin. "Higher spin black holes from CFT." Journal of High Energy Physics 2012.4 (2012): 1-23.
}

\lref\BanadosUE{
  M.~Banados, R.~Canto and S.~Theisen,
  ``The Action for higher spin black holes in three dimensions,''
JHEP {\bf 1207}, 147 (2012).
[arXiv:1204.5105 [hep-th]].
}

\lref\PerezCF{
  A.~Perez, D.~Tempo and R.~Troncoso,
  ``Higher spin gravity in 3D: Black holes, global charges and thermodynamics,''
Phys.\ Lett.\ B {\bf 726}, 444 (2013).
[arXiv:1207.2844 [hep-th]].
}

\lref\CampoleoniHP{
  A.~Campoleoni, S.~Fredenhagen, S.~Pfenninger and S.~Theisen,
  ``Towards metric-like higher-spin gauge theories in three dimensions,''
J.\ Phys.\ A {\bf 46}, 214017 (2013).
[arXiv:1208.1851 [hep-th]].
}

\lref\AmmonWC{
  M.~Ammon, M.~Gutperle, P.~Kraus and E.~Perlmutter,
  ``Black holes in three dimensional higher spin gravity: A review,''
J.\ Phys.\ A {\bf 46}, 214001 (2013).
[arXiv:1208.5182 [hep-th]].
}

\lref\PerlmutterKrausA{
Kraus, Per, and Eric Perlmutter. ``Partition functions of higher spin black holes and their CFT duals." Journal of High Energy Physics 2011.11 (2011): 1-25.
}

\lref\PerlmutterKrausB{
Kraus, Per, and Eric Perlmutter. ``Probing higher spin black holes." Journal of High Energy Physics 2013.2 (2013): 1-38.
}

\lref\ChenPC{
  B.~Chen, J.~Long and Y.~-n.~Wang,
  ``Black holes in Truncated Higher Spin AdS$_3$ Gravity,''
JHEP {\bf 1212}, 052 (2012).
[arXiv:1209.6185 [hep-th]].
}

\lref\DavidIU{
  J.~R.~David, M.~Ferlaino and S.~P.~Kumar,
  ``Thermodynamics of higher spin black holes in 3D,''
JHEP {\bf 1211}, 135 (2012).
[arXiv:1210.0284 [hep-th]].
}

\lref\ChenBA{
  B.~Chen, J.~Long and Y.~-N.~Wang,
  ``Phase Structure of Higher Spin Black Hole,''
JHEP {\bf 1303}, 017 (2013).
[arXiv:1212.6593].
}

\lref\PerezXI{
  A.~Perez, D.~Tempo and R.~Troncoso,
  ``Higher spin black hole entropy in three dimensions,''
JHEP {\bf 1304}, 143 (2013).
[arXiv:1301.0847 [hep-th]].
}

\lref\BoerJottar{
de Boer, Jan, and Juan I. Jottar. ``Thermodynamics of higher spin black holes in AdS3." Journal of High Energy Physics 2014.1 (2014): 1-28.
}

\lref\KrausESI{
  P.~Kraus and T.~Ugajin,
  ``An Entropy Formula for Higher Spin Black Holes via Conical Singularities,''
JHEP {\bf 1305}, 160 (2013).
[arXiv:1302.1583 [hep-th]].
}

\lref\DattaQJA{
  S.~Datta and J.~R.~David,
  ``Black holes in higher spin supergravity,''
JHEP {\bf 1307}, 110 (2013).
[arXiv:1303.1946 [hep-th]].
}

\lref\FerlainoVGA{
  M.~Ferlaino, T.~Hollowood and S.~P.~Kumar,
  ``Asymptotic symmetries and thermodynamics of higher spin black holes in AdS3,''
Phys.\ Rev.\ D {\bf 88}, 066010 (2013).
[arXiv:1305.2011 [hep-th]].
}

\lref\CompereGJA{
  G.~Compère and W.~Song,
  ``${\cal W}$ symmetry and integrability of higher spin black holes,''
JHEP {\bf 1309}, 144 (2013).
[arXiv:1306.0014 [hep-th]].
}

\lref\GaberdielJCA{
  M.~R.~Gaberdiel, K.~Jin and E.~Perlmutter,
  ``Probing higher spin black holes from CFT,''
JHEP {\bf 1310}, 045 (2013).
[arXiv:1307.2221 [hep-th]].
}

\lref\CompereNBA{
  G.~Compère, J.~I.~Jottar and W.~Song,
  ``Observables and Microscopic Entropy of Higher Spin Black Holes,''
JHEP {\bf 1311}, 054 (2013).
[arXiv:1308.2175 [hep-th]].
}

\lref\LiRSA{
  W.~Li, F.~-L.~Lin and C.~-W.~Wang,
  ``Modular Properties of 3D Higher Spin Theory,''
JHEP {\bf 1312}, 094 (2013).
[arXiv:1308.2959 [hep-th]].
}

\lref\HenneauxDRA{
  M.~Henneaux, A.~Perez, D.~Tempo and R.~Troncoso,
  ``Chemical potentials in three-dimensional higher spin anti-de Sitter gravity,''
JHEP {\bf 1312}, 048 (2013).
[arXiv:1309.4362 [hep-th]].
}

\lref\GutperleOXA{
  M.~Gutperle, E.~Hijano and J.~Samani,
  ``Lifshitz black holes in higher spin gravity,''
JHEP {\bf 1404}, 020 (2014).
[arXiv:1310.0837 [hep-th]].
}

\lref\BeccariaDUA{
  M.~Beccaria and G.~Macorini,
  ``On the partition functions of higher spin black holes,''
JHEP {\bf 1312}, 027 (2013).
[arXiv:1310.4410 [hep-th]].
}

\lref\BeccariaYCA{
  M.~Beccaria and G.~Macorini,
  ``Resummation of scalar correlator in higher spin black hole background,''
JHEP {\bf 1402}, 071 (2014).
[arXiv:1311.5450 [hep-th]].
}

\lref\BeccariaGAA{
  M.~Beccaria and G.~Macorini,
  ``Analysis of higher spin black holes with spin-4 chemical potential,''
[arXiv:1312.5599 [hep-th]].
}

\lref\ChowdhuryROA{
  A.~Chowdhury and A.~Saha, ``Phase Structure of Higher Spin Black Holes,''
[arXiv:1312.7017 [hep-th]].
}

\lref\DattaSKA{
  S.~Datta, J.~R.~David, M.~Ferlaino and S.~P.~Kumar,
  ``Higher spin entanglement entropy from CFT,''
[arXiv:1402.0007 [hep-th]].
}

\lref\PerezPYA{
  A.~Perez, D.~Tempo and R.~Troncoso,
[arXiv:1402.1465 [hep-th]].
}

\lref\BunsterMUA{
  C.~Bunster, M.~Henneaux, A.~Perez, D.~Tempo and R.~Troncoso,
  ``Generalized Black Holes in Three-dimensional Spacetime,''
JHEP {\bf 1405}, 031 (2014).
[arXiv:1404.3305 [hep-th]].
}

\lref\DattaUXA{
  S.~Datta, J.~R.~David, M.~Ferlaino and S.~P.~Kumar,
  ``A universal correction to higher spin entanglement entropy,''
[arXiv:1405.0015 [hep-th]].
}

\lref\DattaYPA{
  S.~Datta,
  ``Relative entropy in higher spin holography,''
[arXiv:1406.0520 [hep-th]].
}


\Title{\vbox{\baselineskip14pt
}} {\vbox{\centerline {A new spin on entanglement entropy}}}
\centerline{Eliot Hijano and Per Kraus\foot{eliothijano@physics.ucla.edu, pkraus@ucla.edu}}
\bigskip
\centerline{\it{Department of Physics and Astronomy}}
\centerline{${}$\it{University of California, Los Angeles, CA 90095, USA}}

\baselineskip14pt

\vskip .3in

\centerline{\bf Abstract}
\vskip.2cm

\noindent

We argue that the usual notions of thermodynamic and entanglement entropy have novel analogs in the context of higher spin theories.   In particular,  the Wald and Ryu-Takayanagi formulas have natural higher spin extensions that we work out and study.  On the CFT side, just as standard entanglement entropy in CFT$_2$ can be computed from twist field correlators, we demonstrate that by introducing corresponding operators carrying higher spin charge we can precisely reproduce our results from the bulk.

We also show that the first law for entanglement entropy implies the linearized field equations for the metric and higher spin fields, generalizing recent work on deriving the linearized Einstein equations from the first law.

\Date{June  2014}
\baselineskip13pt

\newsec{Introduction}

\lref\BlancoJOA{
  D.~D.~Blanco, H.~Casini, L.~-Y.~Hung and R.~C.~Myers,
  ``Relative Entropy and Holography,''
JHEP {\bf 1308}, 060 (2013).
[arXiv:1305.3182 [hep-th]].
}

\lref\ProkushkinBQ{
  S.~F.~Prokushkin and M.~A.~Vasiliev,
  ``Higher spin gauge interactions for massive matter fields in 3-D AdS space-time,''
Nucl.\ Phys.\ B {\bf 545}, 385 (1999).
[hep-th/9806236].
}

\lref\GaberdielPZ{
  M.~R.~Gaberdiel and R.~Gopakumar,
  ``An AdS$_3$ Dual for Minimal Model CFTs,''
Phys.\ Rev.\ D {\bf 83}, 066007 (2011).
[arXiv:1011.2986 [hep-th]].
}

\lref\LashkariKOA{
  N.~Lashkari, M.~B.~McDermott and M.~Van Raamsdonk,
  ``Gravitational dynamics from entanglement 'thermodynamics',''
JHEP {\bf 1404}, 195 (2014).
[arXiv:1308.3716 [hep-th]].
}

\lref\SwingleUZA{
  B.~Swingle and M.~Van Raamsdonk,
  ``Universality of Gravity from Entanglement,''
[arXiv:1405.2933 [hep-th]].
}

\lref\AchucarroVZ{
  A.~Achucarro and P.~K.~Townsend,
  ``A Chern-Simons Action for Three-Dimensional anti-De Sitter Supergravity Theories,''
Phys.\ Lett.\ B {\bf 180}, 89 (1986)..
}

\lref\BalasubramanianRE{
  V.~Balasubramanian and P.~Kraus,
  ``A Stress tensor for Anti-de Sitter gravity,''
Commun.\ Math.\ Phys.\  {\bf 208}, 413 (1999).
[hep-th/9902121].
}

Much of the power and appeal of gauge/gravity duality is captured by the statement that it relates  highly fluctuating quantum field theory data to smooth classical geometry, and does so in a precise quantitative way.  The prototype relation is between the thermal entropy of the boundary field theory and the Bekenstein-Hawking entropy of the dual black hole, or its refinement via the Wald entropy \refs{\Wald,\IyerWald}.   This admits a generalization in terms of the Ryu-Takayanagi formula \RyuTakayanagi, which asserts that the entanglement entropy of a subsystem can be computed from the area of a bulk minimal surface that extends to the boundary.   To the extent that understanding quantum gravity means understanding what quantum object has geometry as its classical limit, these relations provide us with  valuable footholds.

It is of obvious interest to ask whether these ideas admit further generalization.  Here we propose such a generalization in the context of the duality between three-dimensional higher spin gravity \ProkushkinBQ\ and certain two-dimensional conformal field theories \GaberdielPZ.   We will argue that there exist new notions of entropy that have the same relation to the higher spin fields that the ordinary entropy has to the metric degrees of freedom.  We will provide several prescriptions for computing the generalized higher spin entropy in a variety of contexts, both in the bulk and in the CFT, and see that an apparently consistent picture emerges.   However, unlike the case for ordinary entropy, the fundamental statistical meaning is unclear at this stage; we lack a first principles formulation in terms of a density matrix, although we anticipate that such a formulation exists.

To motivate our proposal, it is useful to review the ways in which we compute ordinary entropy, and to see how these computations admit natural higher spin generalizations.  First consider the Wald formalism \refs{\Wald,\IyerWald}.  Applied to a stationary black hole solution, one considers the Noether charge corresponding to a Killing vector that vanishes on the horizon (or more accurately on the bifurcation surface).  This leads to a first law relation $T\delta S=\delta E$, where $\delta S$ is a local geometric expression at the horizon, and $\delta E$ is a surface integral at infinity.  As we review, this line of reasoning can also be applied to compute the entanglement entropy of a single interval for small fluctuations around AdS \FaulknerGfC. The analysis is now based on a Killing vector that vanishes on the bulk geodesic connecting the endpoints of the interval.  In higher spin gravity, coordinate transformations are part of the larger higher spin gauge symmetry, and we can therefore ask if there are other gauge transformations that are symmetries of the background and vanish on the appropriate hypersurfaces.  In fact there are, and they lead to first law relations in the same way as in the usual Wald analysis, relating variations at the horizon to those at infinity.  Applied to a single interval on the boundary of AdS, we show that for small fluctuations around AdS this generalized higher spin entropy is equal to the integrated pullback of the higher spin field along the geodesic curve.

The first law relation \BlancoJOA\ for ordinary entanglement entropy is written $\delta S= \delta H_B$, where in the bulk $\delta S$ is the variation of the geodesic length, and $\delta H_B$ is the variation of the ``modular Hamiltonian"
\eqn\Aa{\delta H_B =  - \int_{z_1}^{z_2} \! dz \left({ (z-z_2)(z-z_1) \over z_2-z_1  } \right) T(z)~, }
where the interval is $z\in [z_1,z_2]$ and $T(z)$ is the boundary stress tensor (there is also a  contribution from the anti-holomorphic component of the stress tensor that we have suppressed).  The first law for higher spin entanglement entropy is analogous; for the spin-3 case we write $\delta S_3 = \delta H_{B,3}$.  As noted above, $\delta S_3$ is proportional to the integrated pullback of the spin-3 field, and
\eqn\Aa{\delta H_{B,3} =  - 3\int_{z_1}^{z_2} \! dz \left({ (z-z_2)(z-z_1) \over z_2-z_1 }\right)^2  W(z)~, }
where $W(z)$ is the boundary spin-3 current.

In higher spin gravity, the Ryu-Takayanagi formula is only valid when all higher spin fields are set to zero; when the latter are nonzero there is no preferred metric with which to compute a length. Instead, the correct observable that computes entanglement entropy is a Wilson line in a particular representation of the gauge group \refs{\AmmonCastroIqbal,\BoerJottarB}.   This prescription also turns out to admit a natural version that computes our higher spin entropy.    In particular, the Wilson line that computes $S$ can be thought of as a point particle that carries energy but no higher spin charge.   If we instead take the particle to have vanishing energy but nonzero spin-3 charge, then the Wilson line yields $S_3$ defined in the Wald formalism.    In fact, the Wilson line definition is more general, since it holds for arbitrary backgrounds, not just for stationary black holes or states that are near the vacuum. The charge assignment for the probe particle is not arbitrary, but is fixed by the condition that the particle create a conical singularity in the higher spin field, in the same way that the ordinary probe creates a conical metric singularity.

This conical singularity approach is tightly connected to how one computes entanglement entropy in the CFT via twist fields \CardyCalabrese.  This is based on the replica trick, where one writes $S=-\Tr \rho \log \rho =-[{d\over dn} \Tr \rho^n]_{n=1}$, and then equates $\Tr \rho^n$ with the partition function of the CFT on an n-sheeted Riemann surface.  This partition function can be recast as the correlation function of twist fields in a theory based on n-copies of the original CFT.  The most important property of the twist fields is that they are primaries of scaling dimension $h= {c\over 24}(n-n^{-1}) = {c\over 12}(n-1)+O\big((n-1)^2\big)$.   Given the Wilson line results described above,  it is natural to introduce new ``twist fields" obeying $h=O\big((n-1)^2\big)$ and $w= {c\over 12}(n-1)+O\big((n-1)^2\big)$, where $w$ is the spin-3 charge, and then extract $S_3$ from their correlators by the analogous formula used to obtain $S$.  To test this, we compute the resulting $S_3$ for two cases: for an excited state of the CFT carrying nonzero spin-3 charge, and for the CFT deformed by a spin-3 chemical potential.  In both cases, we find perfect agreement, including numerical factors, with $S_3$ computed in the bulk via the Wilson line or the Wald formalism.    Thus there appears to be a tight logic to our construction.

As just described, we therefore have a recipe for computing $S_3$ in the CFT, even though we don't know precisely what it is that we are computing.  By this we mean that it is not clear at this stage how to define $S_3$ in terms of a reduced density matrix.  For this we need a better understanding of the action of the new twist fields, rather than just knowing their quantum numbers; this is of course an interesting problem for the future.

The first law for entanglement entropy makes a repeated appearance in our work.  Recently, it was shown that by starting from this relation, whose validity holds in the CFT, one can derive the linearized Einstein equations in the bulk \refs{\LashkariKOA,\FaulknerGfC,\SwingleUZA}.  As a slight extension of this idea, we will establish that in the higher spin context the first law can be used to establish the linearized equations for all the higher spin fields and not just the metric.   This is to be expected, since these equations are linked through higher spin gauge invariance.

The remainder of this paper is organized as follows.  In section 2 we discuss the Wald formalism in ordinary gravity, with emphasis on its implementation in the Chern-Simons formulation.  We show how to derive black hole entropy and entanglement entropy from this point of view.  In section 3 we turn to the higher spin theory and show  how the Wald analysis can be generalized to define a new type of entropy. In section 4 we discuss the Wilson line approach to computing generalized higher spin entanglement entropy.  The CFT side is studied in section 5, and in section 6 we show how to obtain the linearized bulk equations from the entanglement first law.     An appendix summarizes our conventions.

\newsec{Wald entropy in the Chern-Simons formulation of 3D gravity}

In this section we first briefly recall the results  of Wald \refs{\Wald}  for defining black hole entropy in an arbitrary diffeomorphism invariant theory of gravity, and then discuss the analogous formalism in the Chern-Simons formulation of 3D gravity \refs{\Witten,\AchucarroVZ}. The latter  formalism will be used in the following sections as it is convenient for introducing higher spin fields in $AdS_3$ \refs{\CampoleoniSSS}.

Wald showed how to define a local geometrical quantity at the horizon of a stationary black hole whose variation $\delta S$ obeys
\eqn\aza{{\kappa \over 2\pi}\delta S = \delta {\cal E}-\Omega_H^{(\mu)}\delta {\cal J}_{(\mu)}}
where $\kappa$ is the surface gravity, ${\cal E}$ and ${\cal J}$ are the canonically defined energy and angular momenta, and $\Omega^{(\mu)}$ are the angular velocities of the horizon.

\subsec{Metric formulation}

We summarize here  the structure of the computation. Starting from a gravitational lagrangian $d$-form $\cal L$, one can compute the symplectic current $\Omega(\delta_{1}\phi,\delta_{2}\phi)$ as follows
\eqn\aa{\Omega(\delta_{1}\phi,\delta_{2}\phi) =\delta_1 \Theta(\delta_2 \phi) -\delta_2 \Theta(\delta_1 \phi)   ,}
where $\Theta(\delta \phi)$ is the ($d-1$) form appearing when one varies the Lagrangian (the total derivative piece),
\eqn\aazb{\delta {\cal L}= E\delta \phi + d \Theta.}
Associated with the diffeomorphism vector field $\xi$ and associated Lie derivative ${\cal L}_\xi$ is the Noether current
\eqn\aaa{J[\xi]=\Theta({\cal L}_\xi \phi)-\xi\cdot{\cal L},}
whose variation is related to the symplectic current in the following way
\eqn\abq{\delta J[\xi]=\Omega(\delta \phi,{\cal L}_{\xi}\phi)+d(\xi\cdot \Theta).}
 On-shell $J[\xi]$ is closed and in fact exact.  Off-shell we can write
\eqn\absz{ J[\xi]=dQ[\xi]+\xi\cdot C,}
where $C$ vanishes when the equations of motion are satisfied.  The expression \absz\ defines the Noether charge $Q[\xi]$.
Now choose  $\xi$  to be a Killing vector that leaves the field configuration invariant, so the symplectic current $\Omega(\delta \phi,{\cal L}_{\xi}\phi)$ vanishes. If we consider an on-shell variation and integrate $\delta J[\xi]-d(\xi \cdot \Theta)$ on a spacelike hypersurface ${\cal C}$ we can write, using \absz,
\eqn\ac{0=\int_{\partial{\cal C}}\left(\delta Q[\xi]-\xi\cdot\Theta(\delta \phi)\right).}
To derive the first law variation we take the hypersurface to have an inner boundary $\Sigma$ at the bifurcation surface of the black hole and an outer boundary at infinity. We further take  $\xi = {\p \over \p t} +\Omega_H^{(\mu)}{\p \over \p \phi^{(\mu)}}$ so that it vanishes at the bifurcation surface.   The boundary term at infinity can then  be identified with $\delta {\cal E}-\Omega_H^{(\mu)}\delta {\cal J}_{(\mu)}$ and we arrive at
\eqn\ad{\delta {\cal E}-\Omega_H^{(\mu)}\delta {\cal J}_{(\mu)}=\int_{\Sigma}\delta Q[\xi].}
This takes the form of the first law under the identification
\eqn\ae{{\kappa \over 2\pi} \delta S =\int_{\Sigma} \delta Q[\xi].}
It remains to integrate \ae\ to obtain $S$.  From  the structure of the Noether charge in an arbitrary diffeomorphism invariant theory of gravity, Iyer and Wald \refs{\IyerWald} thereby obtained their famous result for $S$ in terms of the derivative of the Lagrangian with respect to the Riemann tensor.  For Einstein gravity this of course gives the Bekenstein-Hawking formula $S= A/4G$.

\subsec{Chern-Simons formulation}

We consider Chern-Simons theories with gauge group $G\times G$ and action
\eqn\ba{ I = I_{CS}[A] - I_{CS}[\Ab] }
with
\eqn\bb{I_{CS}[A]={k_{cs}\over 4\pi}\int_M \Tr \left(A\wedge dA +{2\over 3}A\wedge A \wedge A\right).}
The action is invariant under  gauge transformations
\eqn\bc{
\delta_\Lambda A=d\Lambda+[A,\Lambda], \quad\quad
\delta_{\overline{\Lambda}} \bar{A}=d\bar{\Lambda}+[\bar{A},\bar{\Lambda}].
}
The equations of motion imply that the connections are flat
\eqn\baa{
F=dA+A\wedge A =0, \quad\quad \bar{F}=d\bar{A}+\bar{A}\wedge \bar{A}=0.
}
The connections $A$ and $\bar{A}$ are independent and so we can perform the analog of the Wald analysis separately for the two terms in \ba. The symplectic current associated to $A$ is found to be
\eqn\bbz{
\Omega(\delta_{1} A,\delta_{2}A) =-{k_{cs}\over{2\pi}}\Tr\left(\delta_1 A \wedge \delta_2 A\right).
}
Instead of the diffeomorphisms employed in the metric version of Wald's analysis, in the Chern-Simons theory we consider gauge transformations \bc. The symplectic current evaluated for these  gauge transformations is
\eqn\bd{
\Omega(\delta A,\delta_{\Lambda}A) ={k_{cs}\over{2\pi}}\Tr\left(d(\Lambda\delta A)-\Lambda\delta F\right).
}
If $\Lambda$ is a symmetry of the background in the sense that $\delta_\Lambda A=0$, and if both the background and the variation are on-shell, $F=F+\delta F=0$, then integrating \bd\ over a spacelike hypersurface with inner boundary $\Sigma$ and outer boundary at infinity yields
\eqn\be{
{k_{cs}\over{2\pi}}\int_{\infty}\Tr\left(\Lambda\delta A\right)={k_{cs}\over{2\pi}}\int_{\Sigma}\Tr\left(\Lambda\delta A\right).
}
For suitable choice of $\Lambda$, this will be the Chern-Simons analog of the first law variation \ad. The barred connection of course obeys the analogous relation. The  entropy will be identified if we can find an expression $S$ obeying
\eqn\bez{T \delta S ={k_{cs}\over{2\pi}}\int_{\Sigma}\Tr\left(\Lambda\delta A\right)-{k_{cs}\over{2\pi}}\int_{\Sigma}\Tr\left(\Lamb\delta \Ab\right).
}
where $T$ is the temperature.

It is useful to study this in a couple of simple examples.

\subsec{Example 1: Entropy of the BTZ black hole}

The metric of the Euclidean BTZ black hole \refs{\Banados} can be written as
\eqn\daa{
ds^2=d\rho^2+{\cal Q}_2dz^2+\bar{{\cal Q}}_2d\bar{z}^2+\left(  e^{2\rho}+ {\cal Q}_2\bar{{\cal Q}}_2 e^{-2\rho}\right)dzd\bar{z}
}
with
\eqn\daaz{ (z,\zb)\cong (z+2\pi,\zb+2\pi)\cong (z+2\pi \tau, \zb+2\pi \taub)~.}
The charges $\Qc_2$ and $\Qcb_2$ yield the nonvanishing components of the boundary stress tensor \BalasubramanianRE,\foot{this is related to the conventionally defined CFT stress tensor as $T^{\rm grav}={1\over 2\pi}T_{\mu\nu}$.}
\eqn\daay{ T^{\rm grav}_{zz} = {\Qc_2\over 8\pi G}~,\quad T^{\rm grav}_{\zb\zb} ={\Qcb_2 \over 8\pi G}~.}
Smoothness at the Euclidean horizon fixes
\eqn\db{
\tau={{i}\over{\sqrt{4{\cal Q}_2}}},\quad\quad \bar{\tau}=-{{i}\over{\sqrt{4\bar{{\cal Q}}_2}}}.
}
The modular parameter $\tau$ is related to the inverse Hawking temperature $\beta=1/T$ and the angular velocity $\Omega_H$ as
\eqn\dba{ \tau= {i\beta +i\beta\Omega_H \over 2\pi}~,\quad \taub ={-i\beta +i\beta \Omega_H \over 2\pi}~.}
The black hole entropy is read from the area of the horizon at $e^{2\rho_+} = \sqrt{\Qc_2 \Qcb_2}$ as
\eqn\dbb{S = {A\over 4G } = {\pi \over 2G} \left(\sqrt{\Qc_2}+\sqrt{\Qcb_2}\right).}
To obtain this from the Wald analysis one employs the Killing vector (with $z=\phi+it$, $\zb=\phi-it$)
\eqn\dbc{\xi = -i\p_t +\Omega_H \p_\phi =-{2\pi i \over \beta}(\tau \p_z +\taub \p_{\zb})~.}
This can be shown to lead to ${1\over \beta} \delta S = \int_\Sigma \delta Q[\xi]$ in accordance with \ae.

Now we would like to recover this result in the Chern-Simons formulation.  The connections for the BTZ solution are
\eqn\da{\eqalign{
A&=b^{-1}ab+b^{-1}db,\quad \quad a=(L_{1}-{\cal Q}_2 L_{-1}) dz\cr
 \bar{A}&=b\bar{a}b^{-1}+bdb^{-1}, \quad \quad \bar{a}=(L_{-1}-\bar{{\cal Q}}_2 L_{1})d\bar{z}\cr b&=e^{\rho L_0}.
}}
The coordinates have the same periodicity \daaz\ as above. The connections  \da\ represent the BTZ solution for any Chern-Simons gauge group with an SL(2) subgroup, where   the SL(2) generators obey the  algebra
\eqn\dax{ [L_i,L_j]=(i-j)L_{i+j}~.}
We also note that $\Tr ( L_1 L_{-1} ) =-2\Tr(L_0 L_0)$, while other traces of bilinears vanish.

The relations \db\ are imposed by fixing the holonomy of the connection around the thermal circle $S_T$ specified by $(z,\zb) \rightarrow (z+2\pi \tau, \bar{z}+2\pi \bar{\tau})$.  We write
\eqn\dbd{ \oint_{S_T}\! a = 2\pi h~,\quad   h=\tau a_z + \taub a_{\zb}~,}
and then impose
\eqn\dbe{ \Tr [h^2] = - \Tr [L_0 L_{0}]~,}
along with the analogous condition on the barred connection. This ensures that $e^{\oint_{S_T} \!a}$ lies in the center of SL(2), which is the analog of there being no conical singularity at the horizon.

Given the relation between the metric and the connection,
\eqn\dbf{ g_{\mu\nu} ={1\over \Tr (L_0 L_0)}  \Tr (e_\mu e_\nu)~,\quad e_\mu = {1\over 2}(A_\mu -\Ab_\mu)~,}
it is well known that a gauge transformation by
\eqn\dbg{ \Lambda = v\cdot A~,\quad \Lambda = v\cdot \Ab }
acts on the metric as a diffeomorphism by the vector field $v$; i.e.  $\delta g_{\mu\nu} = {\cal L}_v g_{\mu\nu}$.   To obtain the first law variation for the entropy in the Chern-Simons formulation we should take $v=\xi$, where $\xi$ is the same Killing vector \dbc\ employed in the metric formulation.  It is easy to see that such  gauge transformations leave the BTZ connections invariant as required.    From \bez\ we then arrive at
\eqn\dbh{\eqalign{ \delta S &= {{k_{cs} \beta}\over{2\pi}}\int_{\Sigma}\Tr\left(\Lambda\delta A\right)-{{k_{cs}\beta }\over{2\pi}}\int_{\Sigma}\Tr\left(\Lamb\delta \Ab\right) \cr
& = -ik_{cs} \int_0^{2\pi} \! d\phi \Tr \big[h \delta a_\phi\big] + ik_{cs} \int_0^{2\pi} \! d\phi \Tr \big[\hb \delta \ab_\phi\big]~. }}
Note that the integral can be evaluated at any radial location, since the integrand is independent of $\rho$. That is, the inner boundary of $\Sigma$ can be chosen arbitrarily.

Equation \dbh\ is easily integrated to obtain
\eqn\dbi{ S = -2\pi i k_{cs} \Tr [ha_\phi] +2\pi i k_{cs} \Tr [\hb \ab_\phi]~.}
To show this, one uses two facts.  First, one should only allow variations that preserve the thermal holonomy.  This implies that $h$ should only vary by conjugation by an element of the Lie algebra, $\delta h = [h,X]$ for some $X$.  Also, flatness implies that $[h,a_\phi]=[\hb,\ab_\phi]=0$.  Evaluating \dbi\ for the connections \da\ yields
\eqn\dbj{ S= 2\pi k \left(\sqrt{\Qc_2}+\sqrt{\Qcb_2}~\right)}
where we have defined
\eqn\dbja{ k = 2 \Tr(L_0L_0)k_{cs}~.}
$k$ is defined such that $k=k_{cs}$ when the generators correspond to the two-dimensional representation of SL(2).  \dbj\ agrees  with \dbb\ using the identification
\eqn\dbk{  {1\over G}=4k= 8\Tr(L_0 L_0) k_{cs}~.}
The relation \dbk\ follows from comparing the Chern-Simons and Einstein-Hilbert actions.

An alternative form \BoerJottar\ for the entropy can be obtained by diagonalizing the matrices appearing in \dbi.  Note that flatness implies that all the matrices mutually commute and hence can be diagonalized simultaneously.  We can always choose to order the eigenvalues of $h$ and $\overline{h}$ such that (note that this is compatible with \dbe)
\eqn\dbl{ M^{-1} hM =  M^{-1} \hb M = -iL_0~,}
and we then write
\eqn\dbm{ M^{-1} a_\phi M = \lambda_\phi~,\quad  M^{-1} \ab_\phi M = \lamb_\phi~,}
where $\lambda_\phi$ and $\lamb_\phi$ are diagonal.
This yields
\eqn\dbn{ S = 2\pi k_{cs} \Tr [ L_0(\lambda_\phi - \lamb_\phi)]~.}

\subsec{Example 2: Single interval entanglement entropy in Poincar\'e AdS$_3$}

For our next example we consider the holographic computation of entanglement entropy for a single interval in CFT$_2$.  According to the Ryu-Takayanagi formula \refs{\RyuTakayanagi}, for Einstein gravity the entanglement entropy is equal to the (regularized) length of the geodesic connecting the endpoints of the interval,
\eqn\ca{ S={1\over 4G}\int_{\Sigma} \!ds\sqrt{g_{\mu\nu}{{dx^{\mu}}\over{ds}}{{dx^{\nu}}\over{ds}}},}
where $\Sigma$ denotes the geodesic in Poincar\'e AdS$_3$.  It is intructive to derive this result from a Wald type analysis, first in the metric formulation and then in the Chern-Simons formulation.

 The line element of Poincar\'e AdS$_3$ is given by
\eqn\cb{ ds^2={{du^2+dzd\bar{z}}\over{u^2}},}
and the boundary is at $u=0$. The interval is defined to lie in the boundary, with $\bar{z}=z$ and $z\in [z_1, z_2]$.
We will denote this interval as $B(z_1,z_2)$.
The geodesic  anchored at the endpoints of the interval  is given by
\eqn\cba{ u^2+(z-z_1)(z-z_2)=0~,\quad z=\bar{z}.}
In terms of the worldline parameter $s$ we write
\eqn\cbb{ z(s)= \bar{z}(s)= {{z_2+z_1}\over{2}}+ {{z_2-z_1}\over{2}}\tanh(s)~,\quad u(s)={z_2-z_1 \over 2 \cosh(s)}~.}
The tangent vector obeying $g_{\mu\nu}{dx^\mu \over ds}{dx^\nu \over ds}=1$ is
\eqn\cbc{ {dx^\mu \over ds}\p_\mu  ={{z_2-z_1}\over{2\cosh^2(s)}}\left( \p_z +\p_{\bar{z}} -\sinh(s)\p_u\right)~. }
For the Wald analysis we need a Killing vector that vanishes at the geodesic.  Such a Killing vector is
\eqn\cc{ \xi_{B}=-{{2\pi}\over{z_2-z_1}}   \left[  u(z-\bar{z}) \partial_u + \left(  u^2+(z-z_1)(z-z_2) \right)\partial_{z}  + \left( u^2+(\bar{z}-z_1)(\bar{z}-z_2) \right)\partial_{\bar{z}} \right].
}
The surface gravity, computed from $\xi_B^\mu \nabla_\mu \xi_B^\nu =\kappa \xi_B^\nu$ evaluated where $\xi_B \cdot \xi_B=0$, yields $\kappa=2\pi$.   So the corresponding temperature is $T={\kappa\over 2\pi}=1$, which explains how we fixed the normalization of $\xi_B$.

Applying the Wald formulas \ad-\ae\ to this setup yields a first law relation $\delta S_B = \delta H_B$ where $S_B$ is given by \ca\ and $H_B$ is the ``modular Hamiltonian"
\eqn\cn{
H_B =\int_{B(z_1,z_2)} \zeta_B^{\mu} T^{\rm grav}_{\mu\nu} d\sigma^{\nu}~.
}
Here $T^{\rm grav}_{\mu\nu}$ is the boundary stress tensor.  This first law for entanglement entropy can be derived independently in the CFT \refs{\CasiniHuertaMyers}.

 In the Chern Simons formulation, the connections that give rise to the Poincar\'e AdS geometry are
\eqn\cd{\eqalign{A&=b^{-1}ab+b^{-1}db,\quad \quad a=L_1 dz\cr
 \bar{A}&=b\bar{a}b^{-1}+bdb^{-1}, \quad \quad \bar{a}=L_{-1}d\bar{z}\cr b&=e^{L_0 \log{{1}\over{u}}}.
}}
In order to perform a Wald computation we need to use gauge transformations that leave these connections invariant.  A first guess is $\Lambda = \xi_B \cdot A$ and $\Lamb = \xi_B \cdot \Ab$, in analogy with what we had in the BTZ case.  However, one finds that these gauge transformations do not leave the connection \cd\ invariant. More generally, we can take $\Lambda = \xi_B \cdot A+ Y$ and $\Lamb = \xi_B \cdot \Ab+ Y$.  The addition of a common Y to both transformations acts a local Lorentz transformation and has no effect on the metric.  So for a suitable $Y$ we can expect that these transformation will leave the connections invariant, since they all act on the metric as isometries.

The AdS$_3$ metric is invariant under an SL(2) $\times $ SL(2) group of isometries.  The corresponding Killing vectors are
\eqn\cf{\eqalign{
\xi_{L,1} &= \partial_z, \quad \xi_{L,2} =z \partial_z+{{u}\over{2}}\partial_u, \quad   \xi_{L,3} =z^2 \partial_z+zu\partial_u-u^2\partial_{\bar{z}},\cr
\xi_{R,1} &= \partial_{\bar{z}},  \quad \xi_{R,2} =\bar{z} \partial_{\bar{z}}+{{u}\over{2}}\partial_u, \quad \xi_{R,3} =\bar{z}^2 \partial_{\bar{z}}+\bar{z}u\partial_u-u^2\partial_z.
}}
Associated to each Killing vector is a gauge transformation that leaves the connections invariant:
\eqn\cf{
\Lambda_i=\xi_{L,i}\cdot (A-\bar{A}),\quad {\rm and}\quad  \bar{\Lambda}_i=\xi_{R,i}\cdot (A-\bar{A}),
}
where $i=1,2,3$. Now we note that the Killing vector $\xi_B$ can be written
\eqn\cfa{\xi_B = \xi_{B,L}-\xi_{B,R}}
with
\eqn\cfaa{\eqalign{
\xi_{B,L}&={{-2\pi}\over{z_2-z_1}}\left( z_1 z_2 \xi_{L,1} -\left(z_2+z_1\right) \xi_{L,2}+\xi_{L,3}  \right),\cr
\xi_{B,R}&={{-2\pi}\over{z_2-z_1}}\left(z_1 z_2 \xi_{R,1} -\left(z_2+z_1\right) \xi_{R,2}+\xi_{R,3}  \right).
} }
Therefore, the gauge transformation
\eqn\cfb{\eqalign{  \Lambda_B &=\xi_{B,L}\cdot (A-\Ab)=  {{-2\pi} \over {z_2-z_1}} \Big[z_1 z_2 \Lambda_{L,1} -(z_2+z_1) \Lambda_{L,2}+\Lambda_{L,3}\Big]  \cr
\Lamb_B & = \xi_{B,R}\cdot (A-\Ab)=  {{-2\pi} \over {z_2-z_1}} \Big[z_1 z_2\bar{\Lambda}_{R,1} -(z_2+z_1) \bar{\Lambda}_{R,2}+\bar{\Lambda}_{R,3} \Big] }}
leaves the connection invariant, and furthermore acts as an isometry by $\xi_B$ since
\eqn\cfc{\Lambda_B = \xi_B \cdot A +Y~,\quad \Lambda_B = \xi_B \cdot \Ab +Y~,}
with  $Y = \xi_{B,R} \cdot A - \xi_{B,L}\cdot \Ab$.
This gauge transformations is therefore  the appropriate one to use in a Wald computation. We want to show that the relation \be\ is equivalent to the (unbarred part of) the first law $\delta H_B = \delta S_B$, where $S_B$ is given by \ca, and $H_B$ by \cn.  We have
\eqn\cfd{ {k_{cs}\over 2\pi} \int_\Sigma \Tr  [\Lambda_B \delta A]  = {k_{cs}\over 2\pi} \int_\Sigma \Tr [ \xi_{B,L} \cdot (A-\Ab) \delta A ]~.   }
Now we note that when evaluated on the geodesic the Killing vectors $\xi_{B,L}$ and $\xi_{B,R}$ are proportional to the tangent vector along the geodesic,
\eqn\cfe{ \xi_{B,L}^\mu = \xi_{B,R}^\mu = \pi{dx^\mu \over ds}\quad {\rm on~geodesic}~.}
Although the first law relation \be\ holds for any choice of $\Sigma$, if we take $\Sigma$ to be the geodesic then  we can use \cfe\ to integrate  \cfd.
In particular, we then find
\eqn\cff{\eqalign{ {k_{cs}\over 2\pi} \int_\Sigma \Tr  [\Lambda_B \delta A]& = {k_{cs}\over 2} \int_\Sigma \Tr\Big[ {dx\over ds}\cdot (A-\Ab)\delta A\Big] \cr & =
2k_{cs}\int_\Sigma \Tr \Big[ {dx\over ds}\cdot e \delta e \Big] \cr & =
{1\over 4G} \int_\Sigma \! ds \delta \sqrt{g_{\mu\nu} {dx^\mu \over ds}{dx^\nu \over ds}}\cr
&= {\delta A \over 4G}~.   }}

It is also interesting to study the left hand side of equation \be, which is evaluated at the AdS boundary.   Here we consider variations of the connection that preserve the asymptotic AdS$_3$ structure.  We can choose a gauge such that $\delta A= -b^{-1} \delta \Qc_2 L_{-1}bdz$ near the boundary.  Using that $\xi_{B,L}^z =\xi_B^z$ on the boundary, we find
\eqn\cfg{\eqalign{    {k_{cs}\over 2\pi} \int_{B(z_1,z_2)}  \Tr  [\Lambda_B \delta A] & =  -{k_{cs}\over 2\pi} \Tr(L_1 L_{-1}) \int_{B(z_1,z_2)}\! \xi_B^z\delta \Qc_2~dx \cr
&= {1\over 8\pi G} \int_{B(z_1,z_2)}\! \xi_B^z \delta \Qc_2~dx \cr
& = \int_{B(z_1,z_2)}\!\xi_B^z\delta T^{\rm grav}_{zz}~dx\cr
& = \delta H_B~. }}
We have thus  verified the first law for entanglement entropy $\delta S = \delta H_B$.  Of course, the same analysis goes through when we consider a variation of the barred connection.

\newsec{Wald entropy, and its generalizations, in higher spin gravity}

As reviewed in the last section, Wald's analysis takes a Killing vector that vanishes on the bifurcation surface and uses it to construct an entropy that obeys a first law, relating a variation at the horizon to a variation at infinity.   Applied to higher spin gravity, several new issues and possibilities arise.   First, the formalism can be applied to the higher spin black holes constructed in \refs{\GutperleKraus}.   Here the main subtlety involves expressing the result in terms of the right set of charges, as we review in the next subsection.  Second, a new feature is that higher spin gravity enlarges the gauge algebra, and allows for a larger set of gauge transformations on which to base a Wald analysis.  In particular, besides the usual spin-2 gauge transformations employed in the derivation of the thermodynamic entropy, there are additional higher spin transformation that also have the correct properties to lead to first law type variations.   We discuss how these ``higher spin entropies" emerge in several contexts.

\subsec{Thermodynamic entropy of higher spin black holes}

We briefly discuss this case in order to summarize the improved understanding achieved since the original proposal \refs{\GutperleKraus}.\foot{See \refs{\AmmonNK,\CastroFM,\TanTJ,\HartmanGaberdiel,\BanadosUE,\PerezCF,\CampoleoniHP,\AmmonWC,\PerlmutterKrausA,
\PerlmutterKrausB,\ChenPC,\DavidIU,\ChenBA,\PerezXI,\BoerJottar,\KrausESI,\DattaQJA,\FerlainoVGA,\CompereGJA,
\GaberdielJCA,\CompereNBA,\LiRSA,\HenneauxDRA,\GutperleOXA,\BeccariaDUA,\BeccariaYCA,\BeccariaGAA,\ChowdhuryROA}

\refs{\DattaSKA,\BunsterMUA,\DattaUXA,\DattaYPA} for subsequent work.}   The relevant points can all be made in the context of SL(3) $\times $ SL(3) gravity, and so here we restrict to that case.

In \refs{\GutperleKraus} the following connection was considered\foot{We just refer to the unbarred connection, but everything we say carries over to the barred sector.}
\eqn\da{\eqalign{  a_z & =  L_1 - \Qc_2 L_{-1} - \Qc_3 W_{-2} \cr
a_{\zb} & = \mu W_2 + \ldots~.}}
The form of the $\ldots$ terms in $a_{\zb}$ is fixed by demanding flatness of the connection.  In general, one can allow the potential $\mu$ and the currents $\Qc_{2,3}$ to depend on the coordinates $(z,\zb)$.    Doing so, one finds that the flatness conditions take precisely the form of Ward identities for a spin-2 current $\Qc_2$ (i.e. the stress tensor) and spin-3 current $\Qc_3$ in a CFT deformed by a spin-3 potential $\mu$ coupling to the spin-3 current.  Since the Ward identities capture all the information implied by symmetry, this fixes the relation between the quantities in the connection and the currents appearing in a CFT with  $W_3$ symmetry.  For example, the Ward identities can be used to reconstruct uniquely all the correlators of the symmetry currents. This interpretation also passes several other tight consistency checks \refs{\PerlmutterKrausA,\HartmanGaberdiel,\PerlmutterKrausB,\GaberdielJCA}.

These black hole solutions have $\mu$ constant, and the boundary coordinates live on a torus of modular parameter $\tau$.    The relation between the charges $(\Qc_2, \Qc_3)$ and the potentials $(\tau, \mu)$ is fixed by demanding that the holonomy around the thermal circle matches that of the BTZ black hole; we impose \dbe\ along with $\Tr [h^3] = i\Tr[L_0L_0 L_0]$.

In \refs{\GutperleKraus} the entropy was taken to be $S_{\rm hol}=-2\pi ik_{cs} \Tr [ha_z]$ (this way of writing $S$ first appeared in \refs{\BoerJottar, \KrausESI}).   The motivation for this formula is that $\delta S_{\rm hol}$ obeys the same first law variation as the entropy derived from the partition function\foot{We are not keeping track here of the normalization factors in front of the charges, but these can be found in the references cited.}
\eqn\db{ Z = \Tr \left[ e^{4\pi^2 i(\tau \Qc_2+ \alpha \Qc_3)} \right]~,\quad  \alpha = \taub \mu~.  }
Indeed, it was found that the black hole entropy matches the entropy of a CFT in the ensemble defined by \db\ \refs{\HartmanGaberdiel,\PerlmutterKrausA}.  The agreement is robust in that the CFT computations in \refs{\HartmanGaberdiel} only use the symmetry algebra and not fine grained data of the specific CFT under consideration.

However, it was pointed out in \refs{\PerezXI,\BoerJottar} that the black hole entropy computed by canonical methods is not $S_{\rm hol}$ but rather $S_{\rm can} = -2\pi i k_{cs} \Tr[ha_\phi]$, as we saw in arriving at \dbi\ via the Wald formalism.\foot{Note that although we derived \dbi\ in the context of the BTZ black hole, it's easy to see that the logic carries over directly to the higher spin case, with the same result. }  This at first created a puzzle, since $S_{\rm can} \neq S_{\rm hol}$ in the presence of higher spin charge, and yet $S_{\rm hol}$ was found to match the CFT entropy.

To see how this puzzle is resolved it is useful to consider the analogous issues in the context of a Reissner-Nordstrom black hole in Einstein-Maxwell theory.  The time component of the vector potential $A_t$ is usually taken to vanish at the horizon, which is needed in order that $A_\mu$ is a smooth vector field.  At infinity, one then has $A_t =\mu$, and $\mu$ is identified with the chemical potential.   On the other hand, one can take the point of view that black hole solutions should respect the asymptotics of the vacuum, which has $A_t=0$.   Of course,  one can  perform a gauge transformation to set the asymptotic value of $A_t$ in the black hole to zero.\foot{Note that here we are thinking of the Lorentzian black hole solution; in the Euclidean case the holonomy of $A$ around the thermal circle is gauge invariant and nonzero, and so cannot be set to zero by a single valued gauge transformation.}  This gauge transformation is rather inconsequential, in that it just shifts the value of $A_t$ by a constant, leaving everything else unchanged.

In the higher spin case the leading term in $a_{\zb}$ plays the role of a source coupling to the higher spin current on the boundary, as established via the Ward identities. Its presence means that we are adding to the CFT Lagrangian a source term $\mu_3 \Qc_3$.   However in \db\ we are counting up states in the original CFT without such a deformation.  So to compare we should perform a gauge transformation that sets $a_{\zb}=0$, just as was done for $A_t$ in the Reissner-Nordstrom case.  The difference is that because of the more complicated gauge algebra, the gauge transformation acts nontrivially on the solution.  As above, this gauge transformation is admissible in the Lorentzian solution where there is no thermal circle.    The resulting connection, denoted by primes,  is
\eqn\dc{\eqalign{  a'_z & =  L_1 - \Qc'_2 L_{-1} - \Qc'_3 W_{-2} \cr
a'_{\zb} & =0~.}}
The relation between the primed and unprimed charges is fixed by equating the gauge invariant holonomies around the angular circle,
\eqn\dd{  \Tr[(a_\phi)^n] = \Tr[(a'_\phi)^n]~,\quad n=2,3~.}
The main point is then the following: it is not hard to see that $S_{\rm hol}$  has the same functional dependence on $(\Qc_2,\Qc_3)$ as $S_{\rm can}$ has on $(\Qc'_2,\Qc'_3)$.  Therefore, if we take the entropy to be $S_{\rm can}$, but express it in terms of the primed charges, then the agreement with the CFT entropy is maintained.  In particular, in the CFT we continue to use \db\ but with all quantities thought of as primed.  The question of the proper definition of the charges has been discussed in the works \refs{\PerezCF,\PerezXI,\BoerJottar,\CompereGJA,\CompereNBA,\BunsterMUA}; our interpretation above most closely follows \refs{\CompereNBA}.

To summarize, the correct entropy formula is $S_{\rm can}$ as computed from the Wald analysis.  When expressed in terms of variables for which the connection is asymptotically AdS, the result is in agreement with the CFT entropy.

\subsec{Generalized higher spin black hole entropy}

As discussed in section 2.3, the usual thermodynamic entropy follows by working out the Noether charge corresponding to the gauge transformation
\eqn\ea{ \Lambda = \xi \cdot A~,\quad  \xi= -{2\pi i \over \beta}(\tau \p_z + \taub \p_{\zb})~.}
This gauge transformation has two important features.  First, it leaves the black hole solution invariant, and so the Wald formalism will yield a corresponding first law variation.  Second, since $\Lambda$ is proportional to the holonomy $h$, we were able to integrate the first law to arrive at the entropy $S$ in \dbi.

In the higher spin case we have additional possibilities.   In the metric formulation, what is special about the Killing vector $\xi$ is that it vanishes on the bifurcation surface of the black hole. Similarly, in the presence of a spin-3 field, we can look for spin-3 gauge transformations that leave the solution invariant and that vanish on the bifurcation surface.  The corresponding Noether charge will be related to a ``spin-3 entropy" that obeys a first law variation.  In the Chern-Simons formulation this logic can be implemented as follows.

As usual, we consider connections of the form (focussing as always on the unbarred part)
\eqn\eb{ A= b^{-1} a b +b^{-1}db~,\quad b=e^{\rho L_0}~,}
with $a=a_z dz + a_{\zb}d\zb$.  We take $(a_z, a_{\zb})$ to be constant, in which case flatness implies $[a_z,a_{\zb}]=0$, and assume the identification $(z,\zb) \cong (z+2\pi ,\zb +2\pi) \cong (z+2\pi \tau,\zb+2\pi \taub)$.    The connection is left invariant by the gauge transformation
\eqn\ec{ \Lambda_3  = {{3}\over{2\pi}} [\xi \cdot A \xi \cdot A ]|_{tr}~,}
where $|_{\tr}$ refers to the traceless part, so that  $\Lambda_3$ lies in the Lie algebra. We'll see in the next section how such gauge transformations are related to spin-3 transformations that vanish on the bifurcation surface. Note that we have chosen to add a factor of $3/2\pi$ in the definition of $\Lambda_3$. This will be helpful when we compare the results from the Wald formalism and the Wilson line approach.

Repeating the logic leading to \dbi, we now have
\eqn\ed{\eqalign{ T^2\delta S_3 & = {k_{cs}\over 2\pi}\int_\Sigma \Tr(\Lambda_3 \delta A)\cr
& = -3 k_{cs} T^2  \int_0^{2\pi} \! d\phi \Tr [h^2 \delta a_\phi]~.}}
This is integrated to yield
\eqn\ee{ S_3 =-6\pi k_{cs} \Tr[h^2a_\phi]~.}
Note that the factor of $T^2$ accompanying $\delta S_3$ allows the variation to be integrated.

$S_3$ obeys a first law variation of the form
\eqn\ef{ T^2 \delta S_3 = {k_{cs}\over 2\pi} \int_\infty \Tr(\Lambda_3 \delta A)~.}
We will have more to say about the interpretation of the  right hand side in the next section.

Using $W_0 = [L_0 L_0]|_{\tr}$, the same logic that led to \dbn\ now yields
\eqn\eg{ S_3 =6\pi  k_{cs} \Tr[W_0\lambda_\phi]~. }

Now let us see how these consideration lead to an interesting extension of Cardy's formula for the entropy \refs{\Cardy}.   For simplicity we continue to focus on the case of SL(3) gravity.    We now refer to the ordinary thermodynamic entropy as $S_2$, given by $S_2 = 2\pi k_{cs}\Tr[L_0 \lambda_\phi]$.  Being a traceless $3\times 3$ matrix, $\lambda_\phi$ has two independent eigenvalues, and $S_2$ and $S_3$ are two linearly independent combinations of these two eigenvalues.   At the same time we have
\eqn\eh{ a'_\phi = L_1 -\Qc'_2 L_{-1} -\Qc'_3 W_{-2} }
where we are using the primed quantities defined in the previous section. Recall that $a'_\phi$ has the same eigenvalues as $a_\phi$, and hence as $\lambda_\phi$.  By taking traces we have
\eqn\ei{\eqalign{ \Tr[(a'_\phi)^2]& =\Tr[\lambda_\phi^2] =  -2 \Tr(L_1 L_{-1}) \Qc'_2\cr
\Tr[(a'_\phi)^3]& =\Tr[\lambda_\phi^3] =  -3 \Tr(L_1L_1 W_{-2} ) \Qc'_3 }}
Since $\Qc'_2$ and $\Qc'_3$ are given by, respectively, linear combinations of squared or cubed eigenvalues of $\lambda_\phi$, it now follows that there exist relations of the form
\eqn\ej{\eqalign{ \Qc'_2& = c_1 S_2^2 + c_2 S_3^2 \cr
 \Qc'_3& = d_1 S_2^2 S_3 + d_2 S_3^3  }}
for some numerical coefficients $c_{1,2}$ and $d_{1,2}$ which can easily be worked out from the above expressions.     In writing the above we have also used the existence of a charge conjugation symmetry that flips the sign of $\Qc_3$ and $S_3$.   The thermodynamic entropy is obtained by solving these equations to get $S_2= S_2(\Qc'_2,\Qc'_3)$.    The BTZ black hole has $\Qc_3 =S_3=0$, and so the BTZ entropy formula fixes the coefficient $c_1$.  Also, we can choose to set $c_2=1$ by using the freedom to rescale $S_3$.   This leaves the two coefficients $d_{1,2}$.  So just based on general principles, we can say that the entropy formula $S_2(\Qc'_2,\Qc'_3)$ is completely determined up to two numerical parameters.  These two parameters can, for example, be read off from the expansion of $S_2$ to the first two nontrivial order in $\Qc_3$.       It would be interesting to understand from a more general perspective why the simple relations \ej\ hold.  Note that a priori one could modify the right hand side by including an arbitrary function of the dimensionless combination $S_3^2/S_2^3$.     In any case, the system of equations \ej\ seems to be the simplest way to think about the charge dependence of the entropy of a higher spin black hole.

These considerations extend straightforwardly to larger higher spin algebras.  The difference is that now one has more charges, and more coefficients  appear on the right hand of the formulas analogous to \ej.

\subsec{Generalized higher spin entanglement entropy}

In section 2.4 we saw how the entanglement entropy for a single interval could be obtained by applying the Wald formalism.  Here we perform an analogous computation to obtain a generalized entanglement entropy that is based on spin-3 gauge transformations.

We first make some preliminary observations in the metric formulation.
On AdS$_3$ with line element $ds^2 = d\rho^2 +e^{2\rho} dzd\zb$ we consider a symmetric tensor $\varphi_{\alpha\beta\gamma}$ with linearized gauge invariance
\eqn\fa{\delta \varphi_{\alpha\beta\gamma} = \nabla_{(\alpha}\lambda_{\beta \gamma)}~,\quad \lambda^\alpha_\alpha =0~.}
We first look for the general solution of %

\eqn\nqm{  \nabla_{(\alpha}\lambda_{\beta \gamma)}=0~.}
We find  two independent solutions $\lambda^{L}_{\alpha\beta}$ and $\lambda^{R}_{\alpha\beta}$  expressed in terms of functions $f(z)$ and $\fbar(\zb)$ as
\eqn\nqna{\eqalign{ \lambda^{L}_{\rho\rho} &= {1\over 3} \p_z^2 f(z)~,\quad
\lambda^{L}_{\rho z}= {1\over 12} \p_z^3 f(z)~,\quad
\lambda^{L}_{\rho \zb}= -{1\over 2} e^{2\rho} \p_{z} f(z)  \cr
\lambda^{L}_{zz}&= {1\over 24}\p_z^4 f(z)~,\quad
\lambda^{L}_{\zb\zb} = e^{4\rho} f(z) ~,\quad
\lambda^{L}_{z\zb} = -{1\over 12}e^{2\rho} \p_z^2 f(z)  }}
with
\eqn\nqnb{ \p_{z}^5f=0 }
and
\eqn\nqnaa{\eqalign{ \lambda^{R}_{\rho\rho} &=  {1\over 3} \p_{\zb}^2 \fbar(\zb)  ~,\quad
\lambda^{R}_{\rho z}= -{1\over 2} e^{2\rho} \p_{\zb} \fbar(\zb)   ~,\quad
\lambda^{R}_{\rho \zb}= {1\over 12} \p_{\zb}^3 \fbar(\zb)  \cr
\lambda^{R}_{zz}&= e^{4\rho} \fbar(\zb)~,\quad
\lambda^{R}_{\zb\zb} = {1\over 24}\p_{\zb}^4  \fbar(\zb) ~,\quad
\lambda^{R}_{z\zb} = -{1\over 12}e^{2\rho} \p_{\zb}^2 \fbar(\zb)   }}
with
\eqn\nqnba{  \p_{\zb}^5 \fbar =0~. }
The general solution has 10 free parameters. We can look for a transformation that vanishes on the geodesic $(z-z_1)(z-z_2)+u^{2}=0$ with $\zb=z$. If we take
\eqn\nqo{ \eqalign{ f(z)&= \pi^2{{(z-z_1)^2(z-z_2)^2}\over (z_2-z_1)^2}~,\quad
 \fbar(\zb)=\pi^2{{(\bar{z}-z_1)^2(\bar{z}-z_2)^2}\over (z_2-z_1)^2} }}
then the combination $\lambda^{L}_{\alpha\beta} - \lambda^{R}_{\alpha\beta}$ vanishes at the geodesic.

Next, we observe that on the geodesic the gauge transformation is simply related to the tangent vector $v^\mu ={dx^\mu \over ds}$ given in \cbc,
\eqn\fb{\left(\lambda^{L}\right)^{\mu\nu}=\left(\lambda^{R}\right)^{\mu\nu}=\pi^2\left(v^\mu v^\nu -{1\over 3}g^{\mu\nu}\right)~,\quad {\rm on~geodesic}~.}

With this in mind, we now turn to the Chern-Simons formulation and consider the gauge transformation
\eqn\fc{ \Lambda^{L}_3 ={{3}\over{2\pi}} \left(\lambda^{L}\right)^{\mu\nu} (A-\Ab)_\mu (A-\Ab)_\nu~,\quad \Lambda^{R}_3 = {{3}\over{2\pi}} \left(\lambda^{R}\right)^{\mu\nu} (A-\Ab)_\mu (A-\Ab)_\nu~,}
where the normalization is the same as in \ec. Note that these gauge parameters are traceless.  They leave the AdS connection invariant and furthermore acts on $\varphi_{\alpha\beta\gamma}$ as a spin-3 gauge transformation by $\lambda^{L}_{\alpha\beta}-\lambda^{R}_{\alpha\beta}$  which, as we have shown, vanishes on the geodesic.

We now consider the first law variation
\eqn\fd{\eqalign{\delta S_3 & ={k_{cs}\over 2\pi} \int_\Sigma \Tr [\Lambda_3 \delta A] \cr
& ={{3\pi}\over{2}} {k_{cs}\over 2\pi}\int_\Sigma\! ds \Tr \big[(v^\mu v^\nu -{1\over 3} g^{\mu\nu})v^\alpha (A-\Ab)_\mu (A-\Ab)_\nu \delta A_\alpha  \big]   \cr
&={k_{cs}\over 4} \delta \int_\Sigma\! ds \Tr \big[v^\mu v^\nu v^\alpha (A-\Ab)_\mu (A-\Ab)_\nu (A-\Ab)_\alpha \big] \cr
&\quad -{k_{cs}\over 4}\int_\Sigma\! ds \Tr\big[ g^{\mu\nu} (A-\Ab)_\mu (A-\Ab)_\nu v^\alpha \delta A_\alpha \big]       }}
Now, $A$ and $\Ab$ represent the connections for AdS, and we note that
\eqn\fe{ g^{\mu\nu}  (A-\Ab)_\mu (A-\Ab)_\nu =  4L_0^2 -2(L_1 L_{-1}+L_{-1}L_1) \propto C_2 }
where $C_2$ is the SL(2) quadratic Casimir.  The last line in \fd\ therefore vanishes by tracelessness of $\delta A$ and hence
\eqn\ff{ S_3 = {k_{cs}\over 4} \int_\Sigma\! ds \Tr \big[v^\mu v^\nu v^\alpha (A-\Ab)_\mu (A-\Ab)_\nu (A-\Ab)_\alpha \big]~.}
Noting that the spin-3 field is defined (up to a possible normalization factor) as
\eqn\fg{ \varphi_{\mu\nu\alpha} = \Tr[  (A-\Ab)_\mu (A-\Ab)_\nu (A-\Ab)_\alpha  ]}
we see that $S_3$ computes the integral of the   pullback of the spin-3 field to  the geodesic
\eqn\fh{ S_3 ={k_{cs}\over 4} A_3~,\quad A_3 \equiv  \int_\Sigma\!ds ~v^\mu v^\nu v^\alpha   \varphi_{\mu\nu\alpha}~.}
To write this in a reparameterization invariant fashion we should divide the integrand by $g_{\mu\nu}v^\mu v^\nu$, which we have been setting equal to 1.

The first law for $S_3$ states
\eqn\fiz{\delta S_3 = {k_{cs}\over 2\pi} \int_{B(z_1,z_2)} \Tr [\Lambda_3 \delta A]~. }
If we take $\delta A$ to preserve the asymptotic AdS$_3$ structure we can take
\eqn\fj{ \delta A = - e^{-\rho} \delta \Qc_2(z) L_{-1} - e^{-2\rho} \delta \Qc_3(z) W_{-2} ~.}
We then have
\eqn\fl{ {k_{cs}\over 2\pi} \int_{B(z_1,z_2)} \Tr [\Lambda_3 \delta A]={k_{cs}\over {6\pi}}  \int_{B(z_1,z_2)} \! \lambda^{\alpha\beta}\delta \varphi_{\alpha\beta z}dx~.}
Defining the ``modular spin-3 charge"
\eqn\fm{ H_3=  \int_{B(z_1,z_2)} \! \lambda^{\alpha\beta}J_{\alpha\beta z}dx }
with the spin-3 current defined as
\eqn\fma{ J_{zzz} ={k_{cs}\over  {6\pi}}\varphi_{zzz} }
we can write the first law in a form that parallels the first law for entanglement entropy,
\eqn\fn{ \delta S_3 = \delta H_3~.}
In the spin-2 case both sides of this equation have a meaning in the CFT: the left hand side is the entanglement entropy and the right hand side is an integral of the stress tensor.  In the spin-3 case the meaning of the left hand side is not immediately apparent.  What CFT quantity corresponds to the integral of the pullback of the spin-3 field, and what if anything does it have to do with entanglement?

It will be useful to have the explicit expression when the variation is \fj, which is
\eqn\hab{\delta S_3 =\delta H_3=  -12 k_{cs} \int_{z_1}^{z_2}\! dz  \left(  {{(z-z_1)(z-z_2)}\over{z_2-z_1}}   \right)^2 \Qc_3(z)~. }

As should be clear, these considerations are easily extended to include fields of spin larger than 3.

\newsec{Generalized entropy from probe actions}

In \refs{\AmmonCastroIqbal,\BoerJottarB} proposals were made for computing entanglement entropy in the Chern-Simons formulation of 3D higher spin gravity.  These proposals are both based on Wilson lines.   The details differ; here we focus on the proposal in \refs{\AmmonCastroIqbal}, where a particular representation was chosen to compute the usual entanglement entropy.  Our main purpose in this section is to argue that by choosing another representation the result gives the generalized higher spin entropies discussed in the last section.

\subsec{Probe action}

We first review the setup in \refs{\AmmonCastroIqbal}.
We consider a Wilson line in representation $R$ of the gauge group,
\eqn\ga{ W_R(C) =\Tr_R \left[ {\rm P} e^{\int_C A} \right]~,}
where $C$ is a fixed contour that may be open or closed depending on context.\foot{For an open contour we choose boundary conditions at the endpoints rather than taking the trace.}  Thinking of $A$ as a Hamiltonian, the trace over the representation space can be computed by a path integral with an appropriately chosen action.  In the limit that the quadratic Casimir of the representation becomes large, the path integral can be evaluated by saddle point approximation.  This is the case for the representation that is conjectured to yield the entanglement entropy, which has a quadratic Casimir that scales as the square of the central charge which is assumed to be large.

Considering the case of SL(N)$\times $ SL(N), the action depends on the following worldline dynamical variables.   $U$ is an SL(N) group element; $P$ lives in the  SL(N) Lie algebra; and $\lambda_n$ are Lagrange multipliers.  The action is
\eqn\gb{ I[U,P] = \int\! ds \left( \Tr (P U^{-1} D_s U) + \sum_n \lambda_n(\Tr(P^n)-c_n)\right)~.}
Here $\Tr(P^2)=k^{ab} P_a P_b$ where $k^{ab}$ is the Killing metric, and traces of higher powers likewise denote contractions with the corresponding invariant tensors.
The covariant derivative is
\eqn\gc{ D_s U  = {dU \over ds} + A_s U- U\Ab_s~,\quad A_s \equiv A_\mu {dx^\mu\over ds}~.}
For a closed contour all variables are assumed to be single valued around the contour, while for an open contour with endpoints on the AdS boundary we impose $U=1$ at the endpoints (though, as in \refs{\AmmonCastroIqbal}, the rationale for this choice is not well understood).

As usual, we will focus on the simple case of SL(3) $\times$ SL(3), but the generalization is straightforward. In this case we have two Lagrange multipliers $\lambda_2$ and $\lambda_3$ associated with the quadratic and cubic Casimirs.   We take $\Tr P^2$ and $\Tr P^3$ to denote the trace in fundamental representation of the respective powers of $P=P^a T_a$, where $\Tr T_a T_b =\delta_{ab}$.

The equations of motion are
\eqn\gd{ U^{-1}D_s U+2\lambda_2 P+3\lambda_3 (P\times P)=0~,\quad {dP\over ds}+[\Ab_s,P]=0}
where
\eqn\ge{P\times P = h_{abc}T^a P^b P^c~,\quad h_{abc} = \Tr (T_{(a}T_b T_{c)})~,}
together with the constraints implied by the Lagrange multipliers.   The on-shell action works out to be
\eqn\gf{ I = -\int\! ds( 2\lambda_2(s) c_2 +3 \lambda_3(s) c_3)~.}

\subsec{Probe quantum numbers}

In this section we determine the quantum numbers of the probes used to compute entanglement entropy and generalized higher spin entropy. We begin by quickly reviewing the replica trick approach to computing entanglement entropy. We consider a region $B$ of a CFT and construct its reduced density matrix as $\rho_{B}=\Tr_{\bar{B}}\rho$. The entanglement entropy is the Von Neumann entropy given by
\eqn\ka{ S_{EE}=  -\Tr\left( \rho_B \log  \rho_B \right)  ~.  }
For those $\rho$ that can be obtained from a path integral, one can use the replica trick to compute $S_{EE}$. One needs to compute the Renyi entropy
\eqn\kb{ S^{(n)}= {{1}\over{1-n}}\log\Tr \rho^n_B~,  }
and take $S_{EE}=\lim_{n\rightarrow 1}S^{(n)}$, which yields
\eqn\kba{ S_{EE}= -\left( {{d}\over{dn}} \Tr \rho^n_B \right)_{n=1}~.  }
 For integer $n$, one computes $\Tr\rho^n_B$ by considering a path integral on an $n$-sheeted Riemann surface ${\cal R}_n$. This surface is obtained by  sewing copies of the field theory in a periodic fashion. Once an expression has been obtained as a function of $n$, analytically continuing $n\rightarrow 1$ (which in general requires knowledge of the asymptotic behavior in the complex n-plane) gives a result for the entanglement entropy.

A bulk computation of the entanglement entropy using the replica trick consists of finding $AdS_3$ geometries which asymptote to ${\cal R}_n$ \refs{\FaulknerRn}. Because of the periodic sewing, the opening angle around the endpoints of the region $B$ will be $2\pi n$. This means that the geometry will have a conical deficit.  If we solve for the geometry of the bulk in the presence of a Wilson line and demand that it looks like a conical deficit close to the endpoints, we should get a condition that fixes the parameters of the probe (Casimirs). This was done in \refs{\AmmonCastroIqbal} for the probe that computes the entanglement entropy.
There is another way of obtaining these results that does not involve looking directly at the metric. One can demand that the holonomy of the connections around the Wilson line is trivial for integer $n$, making the solution smooth.

It suffices to consider a Wilson line extending radially from the boundary with $z=\zb=$ constant.   The connection $A$ obeys the field equation
\eqn\kc{
{{i k_{cs}}\over{2\pi}} F_{\mu \nu}(x)=-\int ds {{dx^{\mu}}\over{ds}} \epsilon_{\mu\nu\rho} \delta^{(3)}(x-x(s))U^{-1}PU~.
 }
The equation for $\bar{A}$ is similar.   For the radial probe we have, using $\eps_{z\zb \rho}={i\over 2}$,
\eqn\kd{ F_{z\zb} = -{\pi \over k_{cs}} \delta^{(2)}(x-x(s))U^{-1}PU~.}
Stokes' theorem reads
\eqn\kda{\oint A =2i \int\! d^2x F_{z\zb} =-{2\pi i \over k_{cs}} U^{-1}P U~.}
Trivial holonomy is therefore the condition
\eqn\kdb{ e^{{2\pi i \over k_{cs}}P}=1~,}
and so the eigenvalues of $P$ must be integer multiples of $k_{cs}$.

Let $P_0$ denote the diagonal form of $P$.   For SL(3), $P_0$ is a linear combination of $L_0$ and $W_0$.  For the case of ordinary entanglement entropy one would expect the probe to excite only the metric and not the higher spin fields, and this is achieved by taking $P_0 \propto L_0$, which was the case considered in \refs{\AmmonCastroIqbal}.   Noting that $L_0 = {\rm diag}(1,0,-1)$ we have
\eqn\kdc{ P_0 = k_{cs}(n-1)L_0  \quad  {\rm (ordinary~entanglement~entropy~}S)~.}
where we note that for $n=1$ we want the boundary geometry to have a single sheet, which is already present without the probe.  For the case of the higher spin entropy $S_3$ it is natural to expect $P_0\propto W_0$.  Since $W_0={1\over 3}{\rm diag}(1,-2,1)$ we have
\eqn\kdc{ P_0 = 3k_{cs}(n-1)W_0  \quad  {\rm (higher~spin~entanglement~entropy~}S_3)~.}

We can relate the probe quantum numbers to the eigenvalues of $L_0$ and $W_0$ in a highest weight representation.  We denote the eigenvalues of the highest weight state as
\eqn\kdca{ L_0 |{\rm hw}\rangle = h |{\rm hw}\rangle~,\quad W_0 |{\rm hw}\rangle = w|{\rm hw}\rangle~.}
The quadratic and cubic Casimir operators are
\eqn\kdd{C_2 = {1\over 2}h^2 +{3\over 2}w^2+\ldots~,\quad C_3 = {3\over 4}w \left(h^2- w^2\right)+\ldots }
where $\ldots$ denote terms  that are subleading in the large charge limit of interest here.
Equating these to $c_2=\Tr (P^2)$ and $c_3=\Tr(P^3)$ obtained from the probe, we find
\eqn\kde{ \eqalign{ h_2&=2k_{cs}(n-1)~,\quad w_2=0\quad\quad  {\rm (spin-2~probe})\cr
 h_3&=0~,\quad w_3=2k_{cs}(n-1)\quad\quad  {\rm (spin-3~probe}) }}
Note that the spin-2 probe has $c_3=0$, while the spin-3 probe has nonzero values for both $c_2$ and $c_3$.

Since the probe actions will be linear in the probe charges, differentiating with respect to $n$ is equivalent to dropping the factors of $n-1$ in the above.  We henceforth compute the entropy by just evaluating the probe action without the factors of $n-1$ in the charges, since this gives the same result.

\subsec{Probe action for closed loop}

We now discuss the case of a closed probe trajectory, the main application being the case that the probe winds around the horizon of a black hole.
In \refs{\AmmonCastroIqbal} the probe was taken to have $c_3=0$.  For our purposes we need nonzero values for both $c_2$ and $c_3$, as noted above.

The on-shell action can be determined from the gauge holonomy around the closed contour.   We first set $A=\Ab=0$, for which a solution to the probe equations of motion is
\eqn\gg{\eqalign{ U(s)&=U_0(s) = u_0 e^{-2\alpha_2(s)P_0 -3\alpha_3(s) P_0\times P_0}~,\quad {d\alpha_i \over ds}=\lambda_i(s) \cr
P(s)&=P_0~,\quad \Tr P_0^2 =c_2~,\quad \Tr P_0^3 =c_3~.}}
Here $u_0$ is a constant group element, and $P_0$ is a constant Lie algebra element.

Now, the flat connections of interest  can always be written in the form
\eqn\gh{ A = LdL^{-1}~,\quad \Ab= R^{-1} dR~.}
The solution in the presence of these connections is then generated from \gg\ by a gauge transformation by $(L,R)$,
\eqn\gi{U(s)= L(s)U_0(0)R(s)~,\quad P(s) =R^{-1}(s)P_0 R(s)~.}
We write the  connections in the usual form
\eqn\gj{ A=b^{-1}ab+b^{-1}db~,\quad\Ab = b\ab b^{-1}+bdb^{-1}~,\quad b=e^{\rho L_0}.}
We take the probe to move purely in the $\phi$ direction and assume that the connections are $\phi$-independent.   In that case
\eqn\gk{ L(s) = e^{-\rho L_0}e^{-a_\phi \phi(s)}~,\quad R(s) = e^{\ab_\phi \phi(s)}e^{-\rho L_0}~.}
The conditions that $P$ and $U$ are single valued on the contour are now worked out to be
\eqn\gl{\eqalign{ &[P_0,e^{2\pi \ab_\phi}]=0\cr
&e^{-2\Delta \alpha_2 P_0 -3\Delta \alpha_3 P_0 \times P_0} = u_0^{-1} e^{2\pi a_\phi}u_0 e^{-2\pi \ab_\phi}~.}}
Now choose matrices $V$ and $u_0 V$ that diagonalize $a_\phi$ and $\ab_\phi$,
\eqn\gm{\eqalign{ e^{-2\pi \ab_\phi} &= V e^{-2\pi \lamb_\phi}V^{-1}\cr
 e^{2\pi a_\phi} &= (u_0V) e^{2\pi \lambda_\phi}(u_0V)^{-1}~. }}
The equations \gl\ then become
\eqn\gn{\eqalign{ &[V^{-1}P_0 V,e^{2\pi \lambda_\phi}]=0 \cr
&V^{-1}  e^{-2 \Delta \alpha_2 P_0 -3\Delta \alpha_3\Pc_0\times \Pc_0  }V = e^{2\pi (\lambda_\phi-\lamb_\phi)}~.}}
We now take
\eqn\go{ P_0 = V \Pc_0 V^{-1} }
where $\Pc_0$ is diagonal.  Then what remains are the conditions
\eqn\gp{\eqalign{& -2 \Delta \alpha_2 \Pc_0 -3\Delta \alpha_3 \Pc_0 \times \Pc_0=2\pi(\lambda_\phi-\lamb_\phi) \cr
&\Tr [\Pc_0^2]=c_2~,\quad \Tr [\Pc_0^3]=c_3~.  }}
The trace conditions fix $\Pc_0$.  Then by taking the trace of the top equation   against $\Pc_0$, we get
\eqn\gq{\eqalign{ -2 \Delta \alpha_2 c_2 -3\Delta \alpha_3 c_3 = 2\pi \Tr [\Pc_0(\lambda_\phi-\lamb_\phi)] }}
which finally gives the on-shell action as
\eqn\gr{ I =  2\pi \Tr [\Pc_0(\lambda_\phi-\lamb_\phi)]~. }

As worked out in the last subsection, the ordinary spin-2 entropy is obtained by taking $\Pc_0=k_{cs}L_0$, yielding,
\eqn\gs{ S_2 = 2\pi k_{cs} \Tr[L_0 (\lambda_\phi-\lamb_\phi)]~, }
in agreement with \dbn.  Similarly, the higher spin entropy $S_3$ is obtained by taking $\Pc_0 = 3 k_{cs} W_0$, which gives
\eqn\gsa{ S_3 = 6\pi k_{cs} \Tr[W_0 (\lambda_\phi-\lamb_\phi)]~, }
in agreement with \eg.

\subsec{Action for Wilson line with endpoints on boundary}

In this section we wish to make contact with the result \fh\ stating that for linearized fluctuations around AdS, the generalized entropy $S_3$ corresponding to an interval on the boundary is equal to the integral of the pullback of the spin-3 field to the geodesic connecting the endpoints of the interval. Assuming AdS asymptotics, by a gauge transformation we can always bring the unbarred connection to the form
\eqn\ha{ a = \left( L_1 -\Qc_2(z)L_{-1}-\Qc_3(z)W_{-2}\right)dz~.}
We also set $\ab = L_{-1} d\zb$; since we'll work to linear order, we can treat fluctuations in $\ab$ independently from those of $a$, and the analysis is identical.  In \hab\ we obtained the spin-3 entropy as
\eqn\haba{\delta S_3 = -12 k_{cs} \int_{z_1}^{z_2}\! dz  \left(  {{(z-z_1)(z-z_2)}\over{z_2-z_1}}   \right)^2 \Qc_3(z)~. }
The integral is, as we have shown before, proportional to the pullback of the spin three field to the geodesic line.

 We now turn to the probe action, following the approach in \refs{\AmmonCastroIqbal}.  The objective is to show that the answer matches \haba\ .  The connections are written in the pure gauge form
\eqn\hf{ A = LdL^{-1}~,\quad \Ab= R^{-1}dR~.}
As boundary conditions we take $U=1$ at the endpoints of the worldline: $U(s_i)=U(s_f)=1$.  We then define $M$ as
\eqn\hg{M= [R(s_i)L(s_i)][R(s_f)L(s_f)]^{-1}~.}
The diagonal form of $M$ is written $\lambda_M$.  The solution for $P$ is written
\eqn\hh{ P(s) = R^{-1}(s)\Pc_0 R(s),\quad \Pc_0 = {\rm constant}.}
The probe action then works out to be
\eqn\hi{ I = \Tr[ \ln(\lambda_M)\Pc_0]~.}
The computation simplifies for worldlines that end near the boundary.  We take the endpoints to lie at $u=e^{-\rho}=\epsilon$.  For small $\epsilon$,
\eqn\hj{ \lambda_M = {\rm diag}\left( {m_1 \over \eps^4},{m_2\over m_1}, {\eps^4 \over m_2}\right)~,}
where $m_{1,2}$ can be read off from the leading behavior of traces in the fundamental representation as
\eqn\hk{\Tr M = {m_1 \over \eps^4}+O(\eps^{-2})~,\quad (\Tr M)^2 -\Tr(M^2) ={2m_2 \over \eps^4}+O(\eps^{-2})~.}
To compute $S_3$ we take $\Pc_0 =3 k_{cs}  W_0$ as in the previous section.  This gives
\eqn\hka{ S_3 = 3 k_{cs} \ln\left({m_1\over  m_2}\right)~.}
For the connections of interest we have
\eqn\hl{\eqalign{ L&= e^{-\rho L_0}{\rm P} e^{-\int\! a}\cr
R& = e^{L_{-1}\zb}e^{-\rho L_0} }}
where for $L$ we path order along the worldline.    The main quantity we need to compute is therefore
\eqn\hm{{\rm P} e^{-\int a} = {\rm P} e^{-\int_{z_1}^{z_2}  \left( L_1 -\Qc_2(z)L_{-1}-\Qc_3(z) W_{-2}\right)dz }~,}
to linear order in $\Qc_{2,3}$.  The linear term is
\eqn\hn{\eqalign{ {\rm P} e^{-\int \!a}\big|_{\rm lin}  &= e^{-(z_2-z_1) L_1} \int_{z_1}^{z_2}\! dz e^{L_1 z}\big(\Qc_2(z)L_{-1}+\Qc_3(z)W_{-2} \big)  e^{-L_1 z} \cr
& =e^{-(z_2-z_1) L_1}\int_{z_1}^{z_2}\! dz \Big((L_{-1}+2z L_0+z^2L_1)\Qc_2(z)\cr
&\quad\quad\quad\quad\quad\quad\quad\quad +  (W_{-2}+4z W_{-1}+6z^2 W_0+4z^3 W_1+z^4W_2 ) \Qc_3(z)  \Big)~.     }}
This gives
\eqn\ho{\eqalign{M& = \Big( 1- \int_{z_1}^{z_2} \! dz (\eps^2 L_{-1}+2z L_0+{z^2\over \eps^2 } L_1)\Qc_2(z)\cr
&\quad+  (\eps^4 W_{-2}+4z\eps^2 W_{-1}+6z^2 W_0+4{z^3\over \eps^2} W_1+{z^4\over \eps^4} W_2 ) \Qc_3(z) \Big)e^{(z_2-z_1)L_1/\eps^2 }e^{-(z_2-z_1)L_{-1}}~,    }}
where we have set $u(s_f)=u(s_i)=\epsilon$. Evaluating the traces, we now extract $m_{1,2}$ as
\eqn\hp{\eqalign{ m_1& = (z_2-z_1)^4\left(1 -2 \int_{z_1}^{z_2} \! dz {{(z-z_1)(z-z_2)}\over{z_2-z_1}}\Qc_2(z) -2 \int_{z_1}^{z_2} \! dz {{(z-z_1)^2(z-z_2)^2}\over{(z_2-z_1)^2}}\Qc_3(z)\right) \cr
m_2& =  (z_2-z_1)^4\left(1 -2 \int_{z_1}^{z_2} \! dz {{(z-z_1)(z-z_2)}\over{z_2-z_1}}\Qc_2(z) +2 \int_{z_1}^{z_2} \! dz {{(z-z_1)^2(z-z_2)^2}\over{(z_2-z_1)^2}}\Qc_3(z)\right) }}
Plugging into \hka\ and expanding to linear order yields
\eqn\hq{ S_3 = -12 k_{cs} \int_{z_1}^{z_2} \! dz \left({{(z-z_1)(z-z_2)}\over{z_2-z_1}}\right)^2\Qc_3(z) ~.  }
We have thus verified that the probe action indeed reproduces the generalized entanglement entropy computed from the Wald analysis in the Chern Simons formulation.

\newsec{Calculations in the CFT}

It was pointed out in \refs{\CardyCalabrese} that the  partition function on the $n$-sheeted Riemann surface ${\cal R}_n$ may be recast  as the  partition function on the plane of n copies of the original CFT in the presence of twist and anti-twist fields $\sigma$ and $\bar{\sigma}$. The  scaling dimension of these fields is obtained by comparing the expectation value of the stress tensor on ${\cal R}_n$ with the stress tensor in the presence of these twist fields inserted at the endpoints of the interval ($z_1$ and $z_2$). The comparison yields (note that $h_2(\overline{h_2})$ stands for the holomorphic(anti-holomorphic) scaling dimension)
\eqn\la{
h_2(n)=\bar{h}_2(n)={{c}\over{24}}\left(n-{{1}\over{n}}\right)~.
 }
In the case of a CFT with ${\cal W}$ symmetry, these twist fields are primaries associated with a highest weight state whose  $L_0$ eigenvalue is $h_2(n)$. The eigenvalues of the other elements of the Cartan subalgebra are zero. i.e. the twist fields carry vanishing higher spin charge.

Note that expanding \la\ around $n=1$ yields equation \kde, which is the quantum number of the probe that computes entanglement entropy in the bulk. This means that including a  bulk Wilson line anchored at a point on the boundary  is equivalent to inserting a twist field  in the CFT at that point, at least insofar as the charges are concerned.

 In this section we propose that the higher spin entanglement entropies discussed in this work may be computed by evaluating the partition function in the presence of higher spin versions of twist and anti-twist fields associated with a different highest weight state. Namely, for the  case of spin $3$, we will work with quantum numbers that match \kde\ to linear order in $(n-1)$. Note that since for these states, $h$ is of order $(n-1)^2$,  the higher spin entanglement entropy vanishes for a CFT in the vacuum. This makes sense since higher spin entanglement entropies vanish for Poincar\'e AdS. This result will of course change when we consider  the CFT  in an excited state close to the vacuum or in the presence of a small higher spin source. We will perform both of these computations in the following subsections. We will  show that the results agree with bulk calculations.

\subsec{CFT in an excited state}

Let us first consider the ordinary entanglement entropy for an interval but in an excited state close to the vacuum. This will be achieved by studying the following state
\eqn\ma{
\left|\psi\right>=\left|0\right>+\epsilon{\cal O}(y) \left|0\right>~.
 }
Here $\epsilon$ is an infinitesimal parameter and ${\cal O}(y)$ is a local operator  inserted at some point $y$. To first order in $\epsilon$, we have
\eqn\mb{
T(z)=\left<\psi\right|T(z)\left|\psi\right>=\epsilon \left<0\right|T(z){\cal O}(y)\left|0\right>+\epsilon \left<0\right|\left[{\cal O}(y)\right]^{\dagger}T(z)\left|0\right>~.
 }
In order for these two point functions to be nonzero we need the operator ${\cal O}$ to be the stress tensor. We then have
\eqn\mc{\eqalign{
T(z)&=\epsilon \left<0\right|T(z)T(y)\left|0\right>+\epsilon \left<0\right|T(\yb)T(z)\left|0\right>\cr
&=\epsilon {c\over 2}{{1}\over{(z-y)^4}}+\epsilon {c\over 2}{{1}\over{(z-\yb)^4}} ~.
 }}
We henceforth omit writing the second contribution since it just goes along for the ride.  The variation of the modular Hamiltonian is then
\eqn\mca{\eqalign{
\delta H&=\int^{z_2}_{z_1}  dz \xi^{z}_B T_{grav}(z)={1\over{2\pi}}\int^{z_2}_{z_1}  dz \xi^{z}_B T(z)\cr
&=-\epsilon {c\over 2}\int^{z_2}_{z_1}  dz {{(z-z_1)(z-z_2)}\over{\left|z_2-z_1\right|}} {{1}\over{(z-y)^4}}~.
 }}
Performing the integral and invoking the first law yields
\eqn\mcb{
\delta S=-\epsilon {c\over 12} \left({{z_2-z_1}\over{(y-z_1)(y-z_2)}}\right)^2~.
 }
We now show how this result can be obtained in an independent way using the twist field prescription. We need to obtain $\Tr \rho^n$ where $\rho$ is defined by the path integral with the insertion of ${\cal O}(y)=1+\epsilon T(y)$.  To compute $\Tr \rho^n$ via a path integral on ${\cal R}_n$ we clearly need to insert a copy of ${\cal O}$ on each sheet of ${\cal R}_n$.   We then recast this as a path integral of the n-fold replicated theory living on a single sheet.   In this description the operator we insert is the product of ${\cal O}$ operators, one for each replica copy, each sitting at the point $y$.

The moral of the story is that in order to calculate the entanglement entropy we need to consider the correlator of two twist fields and the product of $n$ copies of the operator ${\cal O}$. Given the fact that we are working to linear order in $\epsilon$, we can write
\eqn\me{
\prod^n_{m=1}{\cal O}^{(m)}=\prod^n_{m=1}\left(1+\epsilon T^{(m)}(y)\right) \approx 1+\epsilon T(y)~,
 }
where $T$ is now the total stress tensor of the n-fold replicated theory.
With this we have
\eqn\mf{
\Tr\rho^n=C_n \Big(    \left<\sigma(z_1)\overline{\sigma}(z_2)\right>  +\epsilon \left<\sigma(z_1)\overline{\sigma}(z_2)T(y)\right>           \Big)~.
 }
As the twist fields are primaries, the three point function with the stress tensor is fixed by conformal invariance. The answer is
\eqn\mg{
\Tr\rho^n=C_n \left(   1+\epsilon h_2 \left({{(z_2-z_1)}\over{(y-z_1)(y-z_2)}}\right)^2  \right)  {{1}\over{\left|z_2-z_1\right|^{4h}}}
 }
where $h_2$ is the holomorphic scaling dimension of a twist field.
Taking $h_2$ as in \la\ we obtain
\eqn\mh{
S=-\partial_n\Tr\rho^n|_{n=1}={c\over 3} \log\left|z_2-z_1\right|-\epsilon{c\over{12}}\left({{(z_2-z_1)}\over{(y-z_1)(y-z_2)}}\right)^2~,
 }
which agrees with \mcb. We have used here that $C_1=1$ as $\Tr\rho^n|_{n=1}=1$.  The omitted term in \mc\ would be obtained by including the contribution of $T(\yb)$.

We turn now to the spin-3 case. In order to obtain a nonzero ``modular spin 3 charge" we need to excite the CFT by an operator $W(y)$. We then take the state to be
\eqn\na{
\left|\psi\right>=\left|0\right>+\epsilon W(y) \left|0\right>~.
 }
To first order in $\epsilon$, we have
\eqn\nb{\eqalign{
W(z)&=\left<\psi\right|W(z)\left|\psi\right>=\epsilon \left<0\right|W(z)W(y)\left|0\right>+\epsilon\left<0\right|\left[W(y)\right]^{\dagger}W(z)\left|0\right>\cr
&=\epsilon {5c\over 6}{{1}\over(z-y)^6}+\epsilon {5c\over 6}{{1}\over (z-\yb)^6}~,
 }}
and we henceforth omit the second term.
From the bulk we have shown \hab.    Using $c=24k_{cs}$ along with

\eqn\nd{
 W(z)=4k_{cs}{\cal Q}_3(z)={c\over 6}{\cal Q}_3(z)~,
}
we find
\eqn\ne{\eqalign{
\delta S_3 &= -3 \int^{z_2}_{z_1} dz \left({{(z-z_1)(z-z_2)}\over{z_2-z_1}}\right)^2 W(z)\cr
 &=\epsilon {c\over 12}\left({{z_2-z_1}\over{(y-z_1)(y-z_2)}}\right)^3~.
 }}

We now turn to the twist field prescription. We use primary fields with the quantum numbers in \kde, and we denote them $(\sigma_3,\overline{\sigma}_3)$. We refer to them as twist fields, although it is not clear at present what boundary conditions are being twisted.
We are not starting from first principles  in terms of a density matrix,  but rather going directly to the result in terms of twist operators,  now using the spin-3 current instead of the stress tensor. We will still refer to some kind of spin 3 density matrix $\hat{\rho}$, although the density matrix interpretation is not clear at this stage.

We are inserting a copy of $(1+W(y))$ for each copy of the CFT, so we need to compute the correlator of two twist fields with
\eqn\nf{
\prod^n_{m=1}\left(1+\epsilon W^{(m)}(y)\right) \approx 1+\epsilon W(y)~.
 }
This implies
\eqn\nf{\eqalign{
\Tr\hat{\rho}^n&=\hat{C}_n \left(    \left<\sigma_3(z_1)\overline{\sigma}_3(z_2)\right>  +\epsilon \left<\sigma_3(z_1)\overline{\sigma}_3(z_2)W(y)\right>           \right)\cr
&=\hat{C}_n \left( 1 +\epsilon w_3    \left({{z_2-z_1}\over{(y-z_1)(y-z_2)}}\right)^3        \right){1\over{\left|z_2-z_1\right|^{4h}}}~,
 }}
Using $h_3={\cal O}(n-1)^2$ and $w_3=2k_{cs}(n-1)+{\cal O}(n-1)^2$ we obtain
\eqn\nh{
S=-\partial_n\Tr\hat{\rho}^n|_{n=1}=\epsilon  {c\over 12}\left({{(z_2-z_1)}\over{(y-z_1)(y-z_2)}}\right)^3~.
 }
where we assume $\hat{C}_1 =1$ so that $\Tr \hat{\rho}=1$.  This agrees perfectly with \ne.

\subsec{Entropy in the presence of a source}

In the previous subsection we have considered a CFT in an excited state slightly above the vacuum. The computation in the bulk consisted of turning on some small charges such that the connections are the ones written in \ha. In this subsection we consider turning on a small spin 3 source in the CFT, such that the action is modified by a term ${1\over 2\pi} \int d^2z \mu W$, and computing the higher spin entanglement entropy for an interval.  This corresponds to inserting the term
\eqn\oa{
e^{-{1\over{2\pi}}\int d^2z \mu W}
 }
in any correlator. In the bulk, this corresponds to working with the following connections
\eqn\ob{
A=e^{\rho}L_1 dz +\mu W_2 d\bar{z}+L_0 d\rho~, \quad\quad \bar{A}=e^{\rho}L_{-1} d\bar{z} -L_0 d\rho~.
 }
Let us first investigate the bulk calculation. We will use the Wilson line prescription. Using the new connections, one obtains
\eqn\oba{
M=\left[R(s_i)L(s_i)\right]\left[R(s_f)L(s_f)\right]^{-1}~,
 }
with
\eqn\obb{
L=e^{-\rho L_0}e^{-(z-z_1)L_1+(\bar{z}-\bar{z}_1)\mu W_2}~,\quad\quad R=e^{(\bar{z}-\bar{z}_1)L_{-1}}e^{-\rho L_0}~.
 }
A  procedure similar to the one that led to \hq\ results in the following expression for $S_3$:
\eqn\obc{
S_3=-{{12 k_{cs}\mu}\over{z_2-z_1}}~.
 }

We now deal with the CFT computation. We need to evaluate the correlator between two twist fields with the insertion of \oa. Expanding to linear order in $\mu$ we obtain
\eqn\oc{\eqalign{
\Tr\hat{\rho}^n&=\hat{C}_n\left<\sigma_3(z_1)\overline{\sigma}_3(z_2)e^{-{1\over{2\pi}}\int d^2z \mu W}\right>\cr
&=\hat{C}_n\left<\sigma_3(z_1)\overline{\sigma}_3(z_2)\right>-\hat{C}_n{{\mu}\over{2\pi}}\int d^2 z \left<\sigma_3(z_1)\overline{\sigma}_3(z_2)W(z)\right>~.
 }}
The first term will not contribute to the calculation of $S_3$ as $h_3$ is of ${\cal O}\big((n-1)^2\big)$. The integral in the second term reads
\eqn\od{\eqalign{
\int d^2 z \left<\sigma_3(z_1)\bar{\sigma}_3(z_2)W(z)\right>&=w_3\int d^2 z \left({{z_2-z_1}\over{(z-z_1)(z-z_2)}}\right)^3\left<\sigma_3(z_1)\bar{\sigma}_3(z_2)\right>~.}}
The integral can be defined by cutting out small discs around $z_{1,2}$, integrating, and then taking the disc size to zero.  This is equivalent to the following prescription.  First write
\eqn\oda{ {1\over{(z-z_1)^3(z-z_2)^3}} = \p_{\zb} \left( {\zb \over (z-z_1)^3(z-z_2)^3} \right)-\zb \p_{\zb}  \left( {1\over (z-z_1)^3(z-z_2)^3} \right)  }
Now use Stokes'  theorem
\eqn\odb{ \int_M\! d^2z ( \p_z v^z + \p_{\zb} v^{\zb} ) = \oint_{\p M}  ( v^z d\zb - v^{\zb}dz) }
and
\eqn\odc{ \p_{\zb}  \left({1\over z^{n+1}}\right)=2\pi {(-1)^{n}\over n!} \p_z^{n}\delta^{(2)}(z,\zb)~. }
Integrating \oda, the first term on the right hand side vanishes since there is no pole at infinity, and the second term yields the result
\eqn\odd{\int\! d^2z   {1\over{(z-z_1)^3(z-z_2)^3}} =-12 \pi {\zb_1 -\zb_2 \over (z_1-z_2)^5}~.}
Using $w_3=2k_{cs}(n-1)$, the entropy is then
\eqn\oe{
S_3=-\partial_n\Tr\hat{\rho}^n|_{n=1}=-{{12k_{cs}\mu}\over{z_2-z_1}}~
 }
in agreement with \obc.  Therefore, we have shown with two independent calculations that the generalized higher spin entropy can be calculated as the correlator of two primaries whose quantum numbers match the quantum numbers of the Wilson line probe around $n=1$.

\newsec{Linearized field equations from the first law}

Entanglement entropy obeys a first law that is a generalization of the ordinary first law of thermodynamics. In a CFT with a holographic dual, this law imposes a constraint on the bulk field equations. It was shown in \refs{\FaulknerGfC,\LashkariMcDermottRaamsdonk,\SwingleUZA} that for perturbations around the CFT vacuum state for a ball shaped region, these constraints imply the gravitational linearized equations of motion about pure AdS. Note that one does not obtain the linearized equations of motion for the matter fields living in the bulk. It seems that more than the entanglement first law is needed to derive the full set of linearized on-shell conditions.
In this section we will review these arguments in the Chern-Simons formulation of 2+1 dimensional gravity. We will also show that, in the bulk, there is a natural generalization of the entanglement first law that implies the linearized equations of motion for higher spin fields.

\subsec{SL(2) gravity}

It is useful to re-formulate the ideas in \refs{\FaulknerGfC} in terms of Chern-Simons gauge connections. We will consider a single interval in the CFT, of width $2R$ and centered  at $x_0$.  To parallel the higher dimensional discussion, we will refer to this as a ball $B(x_0,R)$. We consider perturbations around the CFT vacuum.   The dual geometry is pure AdS, and the bulk geodesic anchored at the ends of this region is given in \cba\ which we denote by $\Sigma$.  Obtaining the linearized equations of motion in the bulk means showing $\delta F=0$ for an arbitrary fluctuation.

We start by looking at equation \bd\ for off-shell perturbations of the connections. This means that $\delta F\neq0$. If we choose the gauge parameter $\Lambda_B $ that leaves the background field configuration invariant, the symplectic current must vanish. Integrating this symplectic current over a spacelike surface ${\cal C}$ with inner boundary $\Sigma$ and outer boundary $B(x_0,R)$ yields
\eqn\ia{   {{k_{cs}}\over{2\pi}} \int_{B(x_0,R)} \Tr(\Lambda_B \delta A)- {{k_{cs}}\over{2\pi}} \int_{\Sigma} \Tr(\Lambda_B \delta A)= {{k_{cs}}\over{2\pi}} \int_{{\cal C}} \Tr(\Lambda_B \delta F)~.}
As we discussed in previous sections, the terms on the left hand side of this equation are the variations of the modular hamiltonian and the entanglement entropy under a variation of the connection $A$ leaving $\bar A$ unchanged. Imposing the entanglement first law then implies
\eqn\ib{ \int_{{\cal C}} \Tr(\Lambda_B \delta F)=0~.}
Note that given the choice we have made for ${\cal C}$, this is an integral equation for $F_{ux}$ in the coordinate system specified in \cb. From this integral equation one cannot directly obtain a local constraint, because the parameter $\Lambda_B$ depends on the parameters that specify the ball ($x_0$ and $R$). One way to solve this issue is to make use of the gauge transformations \cf. By construction, these transformations leave the background field configuration invariant, so they don't change $\Lambda_B$. However, they do change $\delta F$. If we denote the gauge transformation by $\Lambda_i$, the new constraint reads
\eqn\ic{ \int_{{\cal C}} \Tr([\Lambda_i , \Lambda_B] \delta F)=0~.}
To say that \ic\ follows from \ib\ is to say that the bulk field equations should be invariant under gauge transformations.
We then have three extra integral constraints apart from the original one. There is a two parameter family of linear combinations of the four constraints whose integrand does not depend on the parameters that spcecify the ball. This constraint is given by
\eqn\id{ \int_{{\cal C}}
\Tr\left[ \left(  {{z^2+\alpha z +\beta }\over{u}}L_1   +u L_{-1} -(\alpha +2z)L_0\right) \delta F \right]=0~
}
for arbitrary parameters $(\alpha,\beta)$.
As this constraint has to be obeyed for every value of  $x_0$ and $R$, we conclude that the integrand has to be zero, obtaining then a two parameter family of local constraints. They imply that the $sl(2)$ part of $\delta F_{ux}$ vanishes,
\eqn\ie{\delta F^{sl(2)}_{ux}=0~.}
In order to obtain the other space time components of the linearized field equations one needs to exploit the invariance of the problem under combinations of $sl(2)$ gauge transformations $\Lambda$ and coordinate transformations $\epsilon^{\mu}$. A combined transformation results in the following change of the linearized field strength
\eqn\iea{\delta F_{\mu \nu}\rightarrow \delta F_{\mu \nu}+\partial_{\mu}\epsilon^{\alpha} \delta F_{\alpha \nu}+\partial_{\nu}\epsilon^{\alpha}\delta F_{\mu \alpha} +\epsilon^{\alpha} \partial_{\alpha} \delta F_{\mu\nu}+[\delta F_{\mu\nu},\Lambda]~. }
So if we choose the following gauge parameters that leave the background field configuration invariant
\eqn\if{\epsilon=u\partial_z ,    \quad \quad  \Lambda=-L_1 , \quad \quad  \bar{\Lambda}=0 ~,}
the new constraint reads
\eqn\ig{   2  \delta F_{ux}      +u\partial_z  \delta F_{ux}   - [\delta F_{ux} , L_1] -2 u \delta F_{xt}   =0 ~,}
where the $sl(2)$ superscript has been omitted. Making use of the result $\delta F_{ux}=0$, one obtains $\delta F_{xt}=0$. Using the gauge parameters
\eqn\ih{\epsilon=z\partial_z ,    \quad \quad  \Lambda=-L_0 , \quad \quad  \bar{\Lambda}=0 ~,}
one obtains also $\delta F_{ut}=0$. The same considerations follow through for the barred connection. We have proven then that
\eqn\ih{\delta F^{sl(2)}_{\mu\nu}=0, \quad \quad \delta \bar{F}^{sl(2)}_{\mu\nu}=0~.}
These equations of motion are equivalent to the linearized Einstein equations and the no torsion constraint.

\subsec{SL(N) gravity}

In SL(N) gravity, the flatness conditions for the Chern-Simons gauge field are the full set of field equations for the metric coupled to higher spin fields.   The results of the previous subsection establish that the first law for entanglement entropy imply the field equations for the SL(2) pure gravity subsector of the higher spin theory.  But since gauge invariance links all the field equations, it will come as no surprise to learn that the entanglement first law in fact implies that the full set of linearized equations are obeyed, as we now show.

We being by writing the pure AdS connections in the usual form
\eqn\ja{A=L^{-1}dL \quad \quad L=e^{z L_1}e^{- L_0 \log u}~.}
The gauge transformation $\Lambda_X=L^{-1}X L$ with $X$ a constant Lie algebra element leaves the connection invariant, so it is a symmetry of pure AdS. Under this gauge transformation the linearized field strength transforms as
\eqn\jb{   \delta F\rightarrow \delta F+[\delta F,L^{-1}X L]  ~.}
From the arguments of the previous section we know that the CFT entanglement first law implies that the $sl(2)$ part of $\delta F$ must vanish. This means that the $sl(2)$ components of equation \jb\  must vanish too. Let's assume for the moment that a $\delta F$ exists that satisfies this condition. It would then be clear that one could never obtain
\eqn\jc{    e^{L^{-1}X L} \delta F e^{-L^{-1}X L} = T ~,}
where $T$ is an $sl(2)$ element. However, in $sl(N)$, for two elements $T_1$ and $T_2$ we can always find a $g$ such that $T_1 =g^{-1}T_2 g$. We must then have that $\delta F^{sl(N)}=0$, establishing that all components of the linearized field strength must vanish. In order to obtain this result we have exploited the fact that higher spin gauge transformations mix the higher spin degrees of freedom with the gravitational ones. The constraint imposed by the entanglement first law then also implies on-shell conditions on the higher spin fields.

It is interesting to note that there is an independent way to show linearized flatness that does not involve the CFT first law. To show this we will work with $SL(3)$ as in previous sections, but the arguments follow through for $sl(N)$.  The ``higher spin entropy" obeys an independent constraint written in equations \fiz\ and \fn. This constraint does not have a clear interpretation in the CFT because the meaning of $S_3$ is not inmediately apparent. However, if we assume that some CFT argument implies that this ``higher spin first law" must be obeyed, there exists an additional constraint on the bulk fields. We start by writing the symplectic current equation for the higher spin transformation \fc\ when the fluctuations of the connections are off-shell. We again choose a spacelike surface ${\cal C}$ with inner boundary $\Sigma$ and outer boundary $B(x_0,R)$
\eqn\jca{   {{k_{cs}}\over{2\pi}} \int_{\infty} \Tr(\Lambda_3 \delta A)- {{k_{cs}}\over{2\pi}} \int_{\Sigma} \Tr(\Lambda_3 \delta A)= {{k_{cs}}\over{2\pi}} \int_{{\cal C}} \Tr(\Lambda_3 \delta F)~.}
Demanding that the variations of the ``higher spin entropy" and the ``modular spin-3 charge"  match results in the following integral constraint on the fluctuations of the fields
\eqn\jd{ \int_{{\cal C}} \Tr(\Lambda_3 \delta F)=0~.}
Note that the gauge parameter $\Lambda_3$ does not involve the generators of the sl(2) subalgebra. The trace in this equation is then selecting the higher spin components of $\delta F$. One can perform higher spin gauge transformations that leave the field configuration invariant. These are given by equations \nqna\ - \nqnba\ . For the connection $A$ there are five different transformations, giving a total of six constraints.
\eqn\jd{ \int_{{\cal C}}\Tr([\Lambda^i_3,\Lambda_3] \delta F)=0,   \quad \quad \Lambda^i_3=\lambda^i_{\mu\nu}(A-\bar{A})^{\mu}(A-\bar{A})^{\nu} ~.}
Similar to the $sl(2)$ case, there is a four parameter family of linear combinations of these six constraints whose integrand does not depend on $x_0$ or $R$. Each of these parameters implies the vanishing of a different $sl(3)$ component of $\delta F$. This immediately implies that $\delta F^{sl(3)}_{ux}=0$. Performing the coordinate plus gauge transformations written in \if\ and \ih, one can obtain the other space time components to conclude that $\delta F^{sl(3)}_{\mu\nu}=0$. Written in the metric formulation, this constraint implies the linearized Fronsdal equations for higher spin fields.

We have concluded that, in the same way that the CFT first law implies a bulk on-shell constraint for the metric, the law concerning the ``higher spin entropy" and the ``modular spin charge" imply on-shell conditions for the higher spin fields.

\vskip .3in

\noindent
{ \bf Acknowledgments}

\vskip .3cm

We wish to thank Eric D'Hoker, Michael Gutperle and Aitor Lewkowycz for useful conversations.
P.K. is supported in part by NSF grant PHY-1313986. E.H. acknowledges support from  `Fundaci\'on La Caixa'.

\appendix{A}{Conventions}
We use the conventions of \refs{\CastroGutperle} for the generators of $SL(N,\Bbb{R})$.\foot{Note that these differ from those of  \refs{\GutperleKraus,\AmmonCastroIqbal}.} We label the $SL(2,\Bbb{R})$ subalgebra with $\{L_0,L_{\pm 1}\}$ and in the $3$-dimensional representation of $SL(N,\Bbb{R})$ they read
\eqn\apa{
L_{1}= -\sqrt{2}\left( \matrix{0&0&0\cr
 1&0&0\cr
0&1&0\cr}\right)~, \quad
L_{0}=\left( \matrix{1&0&0\cr
 0&0&0\cr
0&0&-1\cr}\right)~,\quad
L_{-1}=\sqrt{2}\left( \matrix{0&1&0\cr
 0&0&1\cr
0&0&0\cr}\right)~.
 }
They satisfy the hermicity property $(L_j^{\dagger})=(-1)^{j}L_{-j}$. Given the $SL(2,\Bbb{R})$ subalgebra, an explicit representation for the other generators is
\eqn\apb{
W^{(s)}_{m}=(-1)^{s-m-1} {{(s+m-1)!}\over{(2s-2)!}} \underbrace{[L_{-1},[L_{-1},\cdots,[L_{-1},}_{s-m-1\quad{\rm  terms}}L^{s-1}_{1}]\cdots ]].
 }
In $SL(3,\Bbb{R})$ we have
\eqn\apc{ \eqalign{
W_{1}&= -{1\over{\sqrt{2}}}\left( \matrix{0&0&0\cr
 1&0&0\cr
0&-1&0\cr}\right)~,\quad
W_{0}={1\over 3}\left( \matrix{1&0&0\cr
 0&-2&0\cr
0&0&1\cr}\right)~,\quad
W_{-1}={1\over \sqrt{2}}\left( \matrix{0&1&0\cr
 0&0&-1\cr
0&0&0\cr}\right)~,\cr
W_{2}&=2\left( \matrix{0&0&0\cr
 0&0&0\cr
1&0&0\cr}\right)~, \quad
W_{-2}=2\left( \matrix{0&0&1\cr
 0&0&0\cr
0&0&0\cr}\right)~,
 }}
and some of the nonzero traces are
\eqn\apd{ \eqalign{
\Tr(L_{0}L_{0})&=2 ~,\quad\Tr(L_{1}L_{-1})=-4 ~,\cr
\Tr(W_{0}W_{0})&={2\over 3} ~,\quad\Tr(W_{1}W_{-1})=-1 ~,\quad\Tr(W_{2}W_{-2})=4 ~.
 }}
The commutation relations under this conventions read
\eqn\ape{ \eqalign{
[L_m,L_n]&=(m-n)L_{m+n}~,\cr
[L_m,W_n]&=(2m-n)W_{m+n}~,\cr
[W_m,W_n]&=-{1\over{12}}(m-n)(2m^2+2n^2-mn-8)L_{m+n}~.
 }}
The quadratic and cubic Casimir operators are
\eqn\kdd{C_2 = {1\over 2}L_0^2 +{3\over 2}W_0^2+\ldots~,\quad C_3 = {3\over 4}W_0 \left(L_0^2- W_0^2\right)+\ldots }
where $\ldots $ denote terms with generators of nonzero mode index.

On the CFT side, the above corresponds to the following normalization of the ${\cal W}_3$ algebra
\eqn\apf{ \eqalign{
[L_m,L_n]&=(m-n)L_{m+n}+{c\over{12}}m(m^2-1)\delta_{m+n,0}~,\cr
[L_m,W_n]&=(2m-n)W_{m+n}~,\cr
[W_m,W_n]&=-{{1}\over{12}}(m-n)(2m^2+2n^2-mn-8)L_{m+n}+{8\over c}(m-n)\lambda_{m+n}\cr
&+{{5c}\over{6}}{1\over {5!}}m(m^2-1)(m^2-4)\delta_{m+n,0}~,
 }}
with
\eqn\apg{
\lambda_{m}=\sum_{n}L_nL_{m-n}-{3\over{10}}(m+3)(m+2)L_m~.
 }
The mode expansions of our currents read
\eqn\aph{
T(z)=\sum_n {{L_n}\over{z^{n+2}}}~,\quad W(z)=\sum_n {{W_n}\over{z^{n+3}}}~,
 }
so the OPE's that lead to \apf\ are
\eqn\api{ \eqalign{
T(z)T(w)&\sim  {{c}\over{2}}{{1}\over{(z-w)^4}}+{{2T(w)}\over{(z-w)^2}}+{{\partial_{w}T(w)}\over{(z-w)}}~,\cr
T(z)W(w)&\sim  {{3W(w)}\over{(z-w)^2}}+{{\partial_{w}W(w)}\over{(z-w)}}~,\cr
W(z)W(w)&\sim  {{5c}\over{6}}{{1}\over{(z-w)^6}}+{{5T(w)}\over{(z-w)^4}}+{5\over 2}{{\partial_{w}T(w)}\over{(z-w^3)}}\cr
&+{1\over{(z-w)^2}}\left(5\beta\Lambda(w)+{3\over 4}\partial^2_{w}T(w)\right)+{1\over{(z-w)}}\left({5\over 2}\beta\partial_w\Lambda(w) +{1\over 6}\partial^2_{w}T(w)\right)~,
 }}
where
\eqn\apj{
\beta={16\over{22+5c}}~,\quad \Lambda(w)=:T(w)T(w):-{3\over{10}}\partial^2_{w}T(w)~.
 }
The charges in connections \ha\ are related to the CFT currents in the following way
\eqn\apk{ \eqalign{
T(z)&=4 k_{cs}{\cal Q}_2(z)={c\over 6}{\cal Q}_2(z)~.\cr
W(z)&=4 k_{cs}{\cal Q}_3(z)={c\over 6}{\cal Q}_3(z)~.
 }}
This can be shown by studying gauge transformations that leave the connections \ha\ invariant up to a change in the charges. Matching the variation of the charges with the variation of the CFT currents that can be computed from the structure of the OPE's provides the relations \apk.

\listrefs
\end